\documentclass[epj]{svjour}
\usepackage{latexsym}
\usepackage[T1]{fontenc}
\usepackage[utf8]{inputenc}
\usepackage{amsmath}
\usepackage{amssymb}
\usepackage{graphicx}
\usepackage{grffile}
\usepackage{refstyle}
\usepackage[normalem]{ulem}
\usepackage{rotating}
\usepackage{xcolor}
\usepackage{multirow} 
\newcommand{\ag}{\textcolor{black}}
\begin{document}
\title{Early detection of synchrony in coupled oscillator model}
\subtitle{}
\author{Anupam Ghosh
%
}                     
\offprints{anupamghosh0019@gmail.com} 
%
\institute{Department of Aerospace Engineering, Indian Institute of Technology Madras, Chennai 600036, India}

\date{}
%
\abstract{
\ag{In this paper, we study the applicability of an early warning index while studying the transitions to complete and generalized synchronizations in the coupled oscillator models using an unconventional system parameter and the coupling strength as the required control parameters. The coupled oscillator models are widely used and well-documented for studying various aspects of nature. However, the early warning index used in this paper is an explicit function of the mutual information of the coupled oscillators and reaches two different values before the interacting oscillators yield complete and generalized synchronizations. The transitions to synchrony using the unconventional control parameter are associated with a transition to periodic dynamics of the individual oscillators from their initial chaotic dynamics. Besides, when we use the coupling strength as a control parameter, the interacting oscillator exhibits chaotic dynamics during the synchronizations. Our analysis mainly involves different examples of two low-dimensional oscillators. Finally, we extend our study to a network of interacting oscillators. The applicability of the early detection index is verified in all cases.}
} 

\authorrunning{A. Ghosh}\titlerunning{Comprehending occasional uncoupling induced synchronization}
\maketitle
\section{Introduction}
\label{sec:intro}
The coupled oscillator model is used to understand various real-life examples, including a flock of birds, neural dynamics, insect swarms, and schools of fishes~\cite{winfree01,lakshmanan03,balanov08}. Similar models are also used to investigate atmospheric phenomena like the aerosol-cloud-precipitation system~\cite{feingold13}, solar activity \& El Ni$\tilde{\rm n}$o~\cite{muraki18}, greenhouse gas molecules~\cite{go10}, and ocean-atmosphere dynamics~\cite{miller17}. Various nonlinear phenomena, viz., synchronization, swarming, pattern formation, and amplitude death, have been studied extensively using this coupled oscillator model~\cite{pikovsky01,lakshmanan03,balanov08}. However, in this paper, we focus on one of these nonlinear phenomena: synchronization. 
Synchronization --- a universal phenomenon~\cite{pikovsky01} --- implies coordinated motion of the interacting oscillators and has been studied in coupled dynamical systems for a long time~\cite{pikovsky01,bocc02,strogatz03,ma05,balanov08,strogatz07}. This phenomenon became a topic of intense research for coupled chaotic systems after the pioneering work by Pecora and Carroll~\cite{pc1990}. Subsequent studies support the existence of different kinds of synchronizations in coupled chaotic systems, such as complete synchronization, generalized synchronization, lag synchronization, and phase synchronization~\cite{pecora97,pikovsky01,bocc02,arenas08,balanov08,eroglu17,ghosh18,ghosh18_2,sur20,ghosh20}. Although synchronization is a universal phenomenon, on the contrary, it is not the desired state in few cases, and epilepsy and Parkinson's diseases~\cite{dominguez05,hammond07,rubchinsky12}, the thermoacoustic instability~\cite{lieuwen05,culick06,fisher09,sujith21}, and the Millennium footbridge incident in London~\cite{strogatz05} are such examples. Those psychological diseases in our brain are because of the synchronized oscillations of the neuron oscillators. The thermoacoustic instability, obtained in various combustors, has been modelled as the synchronized state of two mutually coupled oscillators~\cite{sujith21}. In addition, the incident of the Millennium footbridge is due to the synchronization between the oscillation of the bridge and the oscillation of the collective pedestrians~\cite{strogatz05}. Consequently, an early prediction of
synchronization becomes an unavoidable requirement, and we use an \emph{early detection index}~\cite{ghosh22} from the perspective of information theory to detect synchrony early. In other words, \emph{this paper aims to study the applicability of an early warning index while studying the transition to complete and generalized synchronizations from desynchronization in various examples of coupled oscillator models.}
\ag{We briefly overview several other measures reported in the literature to study different kinds of synchronizations. The Kuramoto order parameter is widely used to comprehend the synchrony in interacting phase oscillators~\cite{pikovsky01,balanov08}. The framework of recurrence networks and recurrence plots are useful in studying generalized and phase synchronizations between two variables (or time series)~\cite{romano04,romano05,ghosh20_2}. For two coupled Hamiltonian systems, the measure synchronization~\cite{hampton99} is detected using either the equivalence of energies~\cite{wang03} or the average interacting energy~\cite{sur20}. The enhancement of the mutual information during the transition from desynchronization to synchrony has been adopted widely as an index and used in neuron dynamics~\cite{gupta19}, quantum systems~\cite{ameri15}, and ecology~\cite{wilmer12}. The machine learning technique has recently been used to anticipate synchrony~\cite{fan21}.}
The index we use in this paper is an explicit function of the mutual information of the interacting dynamical systems and is convenient for experimental data and mathematical models. Hence, we adopt this index while studying the transition to synchrony from desynchrony through intermittency, which is also detected in various experiments~\cite{mondal17_chaos,raaj19,sujith21}. First, we vary an \emph{unconventional system parameter} (other than the coupling strength) to study the aforesaid dynamical transition to synchrony as we have more liberty (and it is natural) to change a different system parameter other than the coupling strength in those experiments. Subsequently, for the sake of completeness, we also adopt the coupling strength as a control parameter in our study. Although in the literature on synchronization, the dynamical transition to synchrony is already scrutinized using different examples of interacting dynamical systems~\cite{yu01,nurujjaman06,nurujjaman07,nguyen13,seshadri16}, in this paper, without losing any generality, we adopt the examples of low-dimensional oscillators (Lorenz~\cite{lorenz63}, R\"ossler~\cite{roessler76}, Chen~\cite{chen99}, and dynamo~\cite{mainieri99} oscillators, to be specific) to illustrate the results.
In the first part, we use the bidirectional (or mutual) couplings for the interaction of oscillators and an unconventional control parameter. For coupled Lorenz oscillators, we observe chaotic desynchronization initially; then, the participating oscillators become periodic and are in the complete synchronized state. In our investigation, we detect intermittent complete synchronization at an intermediate value of the control parameter. In the intermittent complete synchronized state, the coupled oscillators yield the synchrony intermittently. Besides, we get similar outputs for the coupled Lorenz and dynamo oscillators: initial chaotic desynchronization leads to the generalized synchronization state through intermittency. The last part involves studying these transitions using the coupling strength as a control parameter. The aforementioned example of coupled Lorenz oscillators is used to study the transition to complete synchrony. However, the R\"ossler driven Lorenz oscillators model is used to study the transition to generalized synchrony. Unlike the first part, the interacting oscillators exhibit chaotic dynamics during the synchronized states. Finally, we extend our study to a network of ten mutually coupled R\"ossler oscillators.
The paper is organized as follows: first, we focus on the general model of mutually coupled oscillators and then discuss the early detection index elaborately (Sec.~\ref{sec:model}). The transition to complete and generalized synchronizations using an unconventional control parameter  and the coupling strength are discussed in sections~\ref{sec:mu} and \ref{sec:alpha}, respectively. Finally, we conclude the major results of this paper in Sec.~\ref{sec:conclusion}. 

\section{General model and early detection index}
\label{sec:model}
In this paper, we choose the oscillators, which are autonomous (i.e., the corresponding equations of motions are explicitly time-independent) and three-dimensional in phase space. Let us consider that the vectors $\mathbf{x}_1 (t) := (x_1, {y}_1, {z}_1)$ and $\mathbf{x}_2(t):= ({x}_2, {y}_2, {z}_2)$ represent the respective phase space variables of the interacting oscillators, then the general equations of motions of the mutually coupled oscillators are given by: 
\begin{subequations}
	\label{eq:1}
	\begin{eqnarray}
		\frac{d{\mathbf{x}}_1}{dt} &=& \mathbf{F(x}_1, \mu_1) + \alpha \cdot \mathsf{C} \cdot (\mathbf{x}_2 - \mathbf{x}_1),\label{eq:1a}\\
		\frac{d{\mathbf{x}}_2}{dt} & =& \mathbf{G(x}_2, \mu_2) + \alpha \cdot \mathsf{C} \cdot  (\mathbf{x}_1 - \mathbf{x}_2),\label{eq:1b}
	\end{eqnarray}
\end{subequations}
where $\mathbf{F}(\cdot)$ and $\mathbf{G}(\cdot)$ are the functional forms of the coupled oscillators.  The scalar, $\alpha$, measures coupling strength between $\mathbf{x}_1 (t)$ and $\mathbf{x}_2(t)$. Lastly, $\textsf{C}$ is the coupling matrix, and for the example in hand, $\textsf{C}$ has dimension $3 \times 3$. In Eq.~\ref{eq:1}, we vary the system parameter $\mu_i$ ($i = 1, 2$) and coupling strength $\alpha$ to study the dynamical transition to synchrony. 
Thus, we have armed with the general model of coupled oscillators in which we want to study the transition from desynchrony to synchrony. Now, we are interested in defining the early detection index ($R$) that anticipates synchrony. In order to calculate $R$, we need two variables (or time series), and each variable is corresponding to each interacting system (or oscillator). Let us consider two variables $X = \{ x^j\}_{j = 1}^{N}$ and $Y = \{ y^j\}_{j = 1}^{N}$, then the early detection index between $X$ and $Y$ is defined as:
\begin{equation}
	\label{eq:order}
	R(X, Y) := \frac{H(X) + H(Y)}{2} - I(X;Y),
\end{equation}
where $H(X)$ and $H(Y)$ are the Shannon entropy~\cite{cover06} of the variables $X$ and $Y$, respectively. $I(X;Y)$ is the mutual information between $X$ and $Y$~\cite{cover06}. If $p(x^1), p(x^2), \cdots, p(x^N)$ are the probabilities corresponding to $x^1, x^2, \cdots, x^N$ of the variable $X$, then the Shannon entropy $H(X)$ of the variable $X$ is given by: 
\begin{equation}
	\label{eq:shanon}
	H(X) = - \sum_{j=1}^{N} p(x^j) \log_{2} \left(p(x^j) \right).
\end{equation}
In Eq.~\ref{eq:shanon}, $H(X)$ has unit in bits as we choose base $2$ in the logarithmic function. Similarly, we can define the Shannon entropy $H(Y)$ of the variable $Y$. Going further, mutual information $I(X;Y)$ --- which quantifies the mutual dependence between $X$ and $Y$ --- is defined as:
\begin{equation}
	\label{eq:mi}
	I(X; Y) = H(X) + H(Y) - H(X,Y),
\end{equation}
where $H(X,Y)$ is the joint entropy of the variables $X$ and $Y$, and mathematically, we can define $H(X,Y)$ as follows:
\begin{equation}
	\label{eq:joint_entropy}
	H(X, Y) = - \sum_{j,k=1}^{N} p(x^j, y^k) \log_{2} \left(p(x^j, y^k) \right).
\end{equation}
Furthermore, we may redefine $R$ as follows:
\begin{equation}
	\label{eq:order_para}
	R(X, Y) = \frac{H(X,Y) - I(X;Y)}{2}.
\end{equation}
To this end, we mention that the early detection index is the difference between joint entropy and mutual information. By construction, $R$ is a non-negative real number, and it is symmetric in $X$ and $Y$. $R$ reduces a small number as $X$ and $Y$ yield synchrony. To be more explicit, the early detection index decreases as the interacting systems reach synchrony from the desynchronization state. However, to calculate $R$, we may adopt the $x$-coordinates of Eq.~\ref{eq:1} (i.e., variables $x_1(t)$ and $x_2(t)$). Also, we remove the initial $10\%$ data of variables $x_1(t)$ and $x_2(t)$ as transient while calculate $R$ at a fixed value of the control parameter. One may choose the $y$ or $z$-coordinates also; however, we end up getting the same conclusions in all cases.    
\ag{In this paper, we study two different kinds of synchronization obtained in coupled oscillator models: complete and generalized synchronizations~\cite{pecora97}. Both kinds of synchronization imply coherence in both phases and amplitudes of the interacting oscillators. Nonetheless, unlike complete synchrony, generalized synchronization is detected in coupled non-identical oscillators. In order to understand the difference more clearly, we consider the variables $\mathbf{x_1}$ and $\mathbf{x_2}$ of Eq.~\ref{eq:1}; then complete synchronization requires $\mathbf{x_1} = \mathbf{x_2}$. Besides, for generalized synchronization, we have $\mathbf{x_1} = \phi (\mathbf{x_2})$, where $\phi$ is a functional that connects the phase space variables of one oscillator with the other. Thus, it is easy to identify that $\phi$ is the identity for complete synchronization, whereas it is hard to find out $\phi$ for generalized synchronization. To this end, the early detection index $R$ reaches a small non-zero number during the generalized synchronization state as the generalized synchrony is observed in coupled non-identical oscillators. On the contrary, $R$ reaches zero in the case of complete synchrony.}
Earlier studies have used mutual information as a measure to study synchronization in many
cases, such as quantum systems~\cite{ameri15}, electrophysiological data~\cite{wilmer12}, and neuron dynamics~\cite{gupta19}. It has been reported that the transition from desynchrony to synchrony is associated with an enhancement	of mutual information. This enhancement is detected for both generalized and complete synchronizations. It is not possible to distinguish between complete and generalized synchrony after calculating mutual information between the signals, as mutual information reach arbitrary high values in both cases. The early detection index ($R$) we use here overcomes the discussed problem~\cite{ghosh22}. During the complete synchrony, $R$ reaches zero. Besides, $R$ reaches a small non-zero value during the generalized synchrony.
In Sec.~\ref{sec:result}, we study the transition to two different kinds of synchrony using various examples of coupled oscillators. We verify the applicability of the early detection index in all cases. Note that the fourth-order Runge--Kutta method has been used in this paper to integrate Eq.~\ref{eq:1} numerically. The maximum time has been taken as $1000$, with the smallest time step $0.001$.
\section{Results}
\label{sec:result}
In our analysis, we first adopt the system parameter $\mu_i$ as the required control parameter motivating from the experimental studies (Sec.~\ref{sec:mu}). Furthermore, we choose the coupling strength ($\alpha$) as the control parameter in Sec.~\ref{sec:alpha}. The examples of the low-dimensional oscillators, viz., Lorenz~\cite{lorenz63}, R\"ossler~\cite{roessler76}, Chen~\cite{chen99}, and dynamo~\cite{mainieri99}, have adopted in this paper.
\subsection{Using an unconventional system parameter as a control parameter}
\label{sec:mu}
\subsubsection{Mutually coupled Lorenz}
\label{sec:cs}
\begin{figure}[h]
	\hspace*{1 mm}
	\includegraphics[width=45cm,height=3.5cm, keepaspectratio]{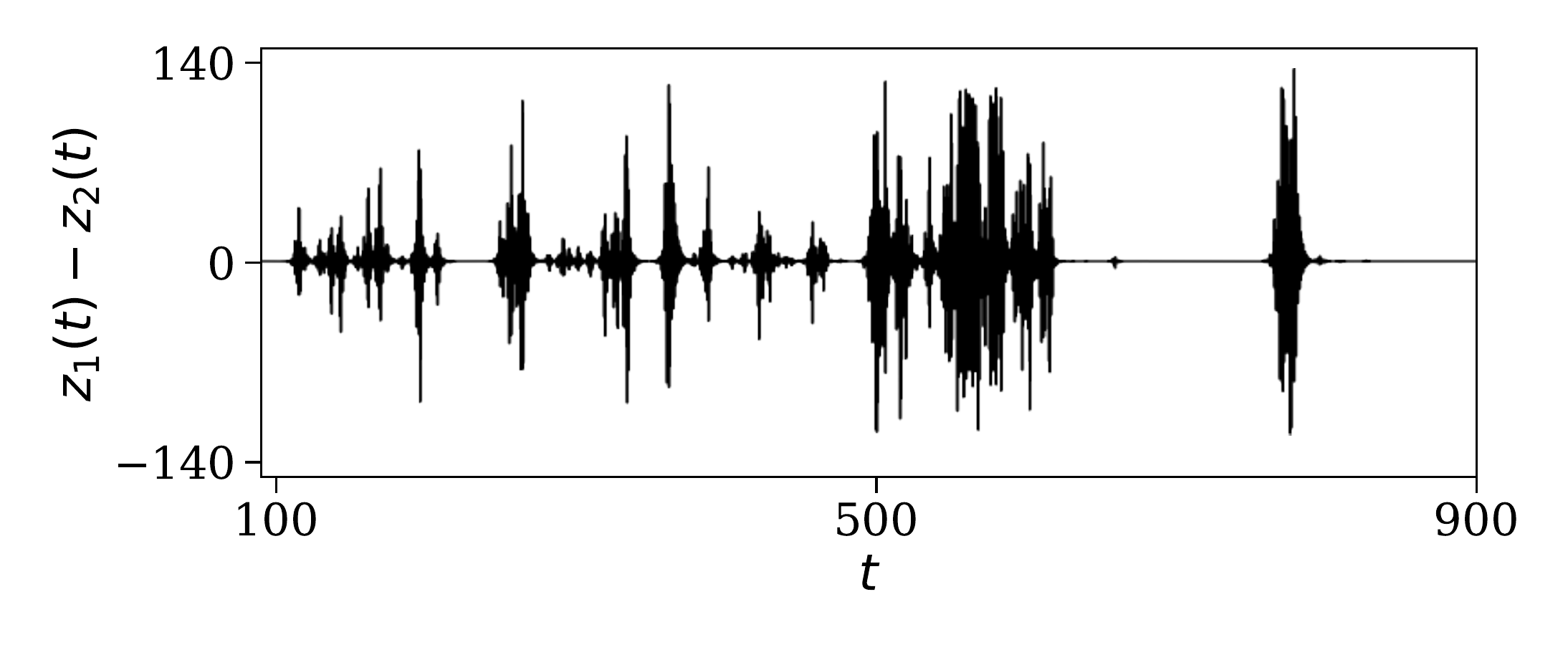}
	\caption{\textbf{Intermittent complete synchronization at $r= 142.9$:} Intermittent complete synchronization is detected at $r = 142.9$. The difference between the $z$-coordinates of the coupled Lorenz oscillators intermittently converge to each other.}
	\label{fig:lorenz_inter}
\end{figure}

Now, we concentrate on the first example: bidirectionally coupled Lorenz oscillators~\cite{lorenz63}. For this coupled Lorenz oscillators, $\mathbf{F}(\cdot)$ and $\mathbf{G}(\cdot)$, in Eq.~\ref{eq:1}, are identical. The explicit form of the interacting oscillators are given by:
\begin{subequations}
	\label{eq:lorenz}
	\begin{eqnarray}
		\frac{d{x}_i}{dt} &=& 10(y_i-x_i),\label{eq:4a}\\
		\frac{d{y}_i}{dt} &=& x_i(r-z_i)-y_i,\label{eq:4b}\\
		\frac{d{z}_i}{dt} &=& x_i y_i - \frac{8}{3}z_i + \alpha (z_j - z_i),\label{eq:4c}
	\end{eqnarray}
\end{subequations}
where $i = 1,2$ and $j = 1,2$ with $j \neq i$. Here, we have fixed the coupling strength parameter at $\alpha = 0.65$ as it is recommended to choose the coupling strength smaller so that the coupling strength parameter perturbs the coupled systems' intrinsic dynamical properties~\cite{pikovsky01}. The initial condition to integrate Eq.~\ref{eq:lorenz} is adopted as $(-0.1, 0, 1.01, 0.01, -0.01, 1)$. We focus on the transition from desynchrony to complete synchrony between the oscillators after increasing the control parameter $r$ monotonically. Note that for the example in hand, to keep consistent with Eq.~\ref{eq:1}, the coupling matrix $\textsf{C}$ has only one non-zero element, i.e., $\textsf{C} = \text{diag}(0,0,1)$, and the control parameter is $\mu_1 = \mu_2 = r$. Furthermore, if the coupled oscillators (Eq.~\ref{eq:lorenz}) lead to the following condition:
\begin{equation}
	\label{eq:5}
	e(t):=||\mathbf{x}_1(t) - \mathbf{x}_2(t)|| \rightarrow 0 \,\,\, {\, \rm as \, } \,\,\, t \rightarrow \infty,
\end{equation}
then $\mathbf{x}_1 (t)$ and $\mathbf{x}_2(t)$ are in the identical (or complete) synchronization state, where $||\cdot||$ represents the standard Euclidean distance.
\begin{figure}[htbp!]
	\hspace*{10 mm}
	\includegraphics[width=55cm,height=8.2cm, keepaspectratio]{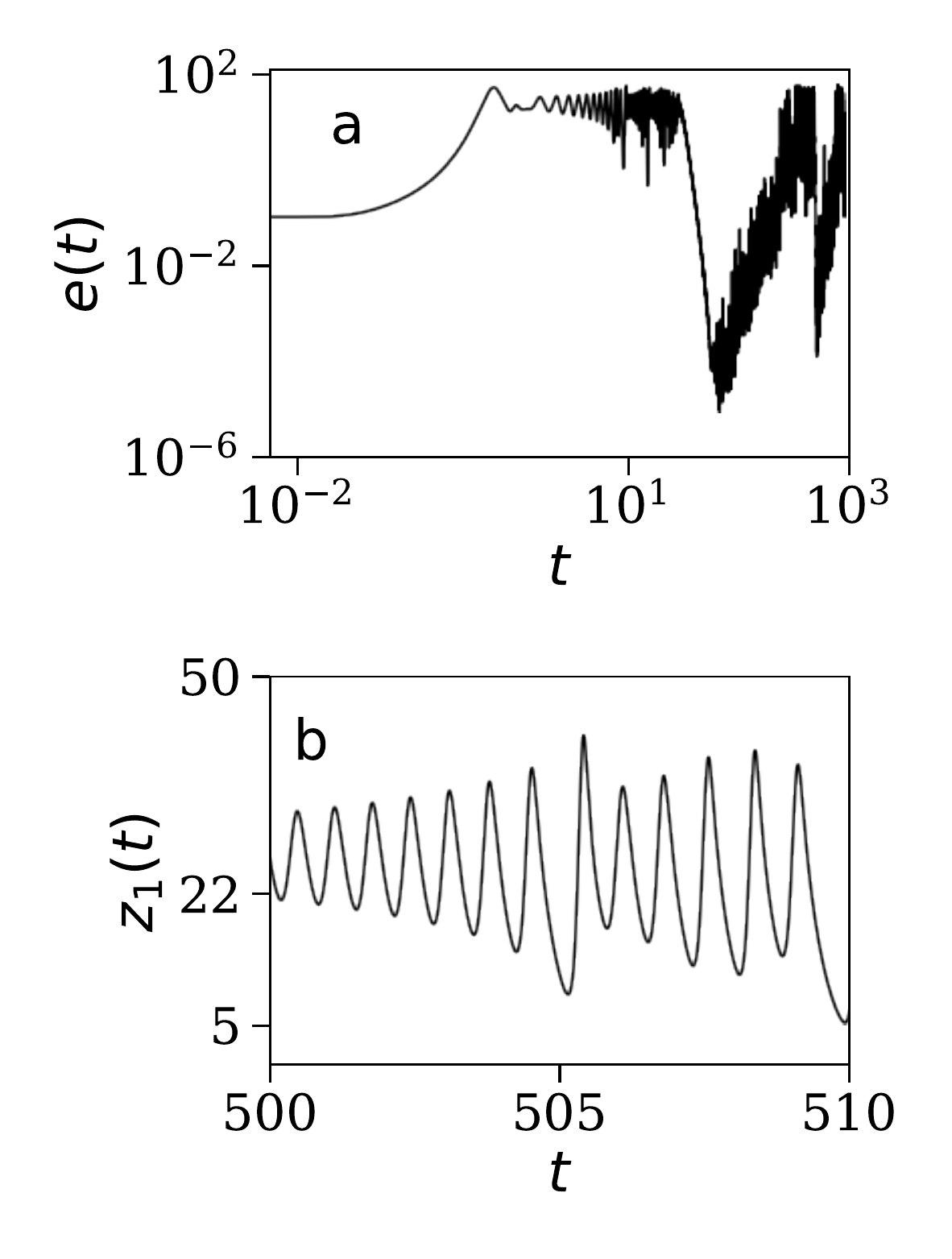}
	\caption{\textbf{Desynchronized state and chaotic dynamics at $r = 28$:} In subplot~(a), the Euclidean norm $e(t)$ (defined in Eq.~\ref{eq:5}) is plotted with the evolution time $t$ for mutually coupled Lorenz oscillators (Eq.~\ref{eq:lorenz}). The norm $e(t)$ does not lead to zero at higher values of $t$ indicating the desynchronization between $\mathbf{x}_1(t)$ and $\mathbf{x}_2(t)$ at $r = 28$. The $z$-component of the first Lorenz oscillator is plotted with $t$ in subplot~(b). It depicts the aperiodic nature of the time series $z_1(t)$.}
	\label{fig:lorenz_desyn}
\end{figure}
We start with $r = 28$, at which the isolated Lorenz oscillator manifests its inherent chaotic nature~\cite{sprott03}. The corresponding outputs at $r = 28$ are depicted in Fig.~\ref{fig:lorenz_desyn}. Since the Euclidean norm $e(t)$ between $\mathbf{x}_1 (t)$ and $\mathbf{x}_2(t)$, in Fig.~\ref{fig:lorenz_desyn}a, does not decrease to zero with the increase in evolution time $t$, desynchronized state is observed. In Fig.~\ref{fig:lorenz_desyn}b, the time series $z_1(t)$ is aperiodic, and the amplitude of oscillations varies within the approximate range $[4.94,47.24]$. We further increase $r$, and at $r = 142.9$ the intermittent complete synchronization is detected (Fig.~\ref{fig:lorenz_inter}). 

\begin{figure}[h]
	\hspace*{10 mm}
	\includegraphics[width=55cm,height=8.2cm, keepaspectratio]{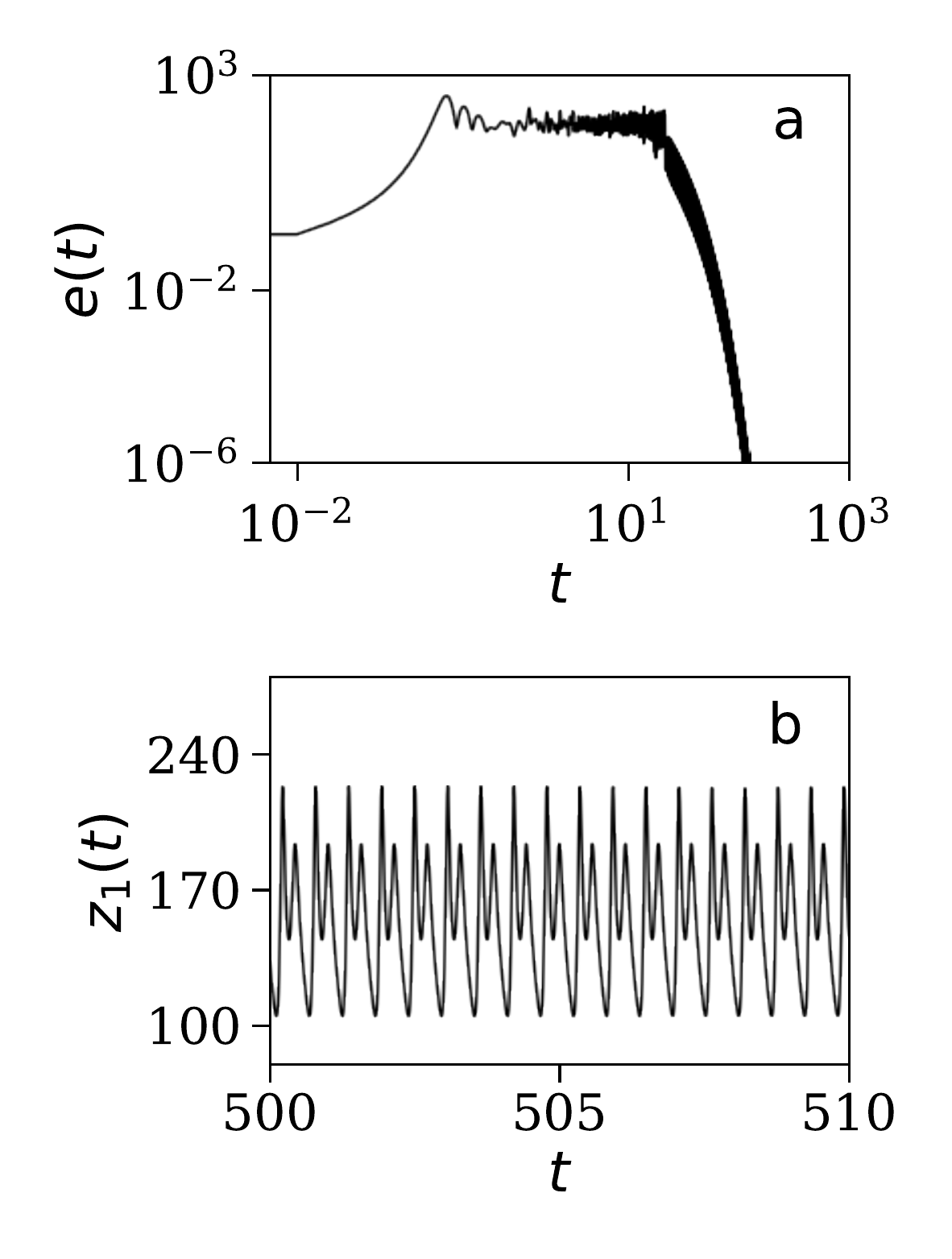}
	\caption{\textbf{Complete synchronized state and periodic dynamics at $r = 164$:} In subplot~(a), the Euclidean norm $e(t)$ leads to zero at higher values of $t$ indicating the synchronization between $\mathbf{x}_1(t)$ and $\mathbf{x}_2(t)$ in Eq.~\ref{eq:lorenz} at $r = 164$. The $z$-component of first Lorenz oscillator is plotted with $t$ in subplot~(b), and it clearly represents the periodic variation of $z_1(t)$ in time $t$.}
	\label{fig:lorenz_cs}
\end{figure}

\ag{The variation of $z_1(t) - z_2(t)$ with $t$, as depicted in Fig.~\ref{fig:lorenz_inter}, implies that occasionally both the time series $z_1(t)$ and $z_2(t)$ overlap and separate from each other. In other words, the coupled Lorenz oscillators intermittently yield the complete synchronized state at $r = 142.9$. Hence, we call this state the intermittent complete synchronization. One can also plot $x_1(t) - x_2(t)$ or $y_1(t) - y_2(t)$ of Eq.~\ref{eq:lorenz} with time $t$ to confirm this intermittent complete synchronization state.} Finally, at $r = 164$, both the oscillators become periodic and lead to synchronized state (Fig.~\ref{fig:lorenz_cs}). In Fig.~\ref{fig:lorenz_cs}a, the Euclidean norm $e(t)$ vanishes with the increase in $t$ indicating the aforesaid synchronization. In addition, Fig.~\ref{fig:lorenz_cs}b refers to the periodic behavior of $z_1(t)$; the amplitude of oscillations increases, and it varies within the approximate range $[105.07, 223.23]$. 
Therefore, to sum up, as the parameter $r$ increases from $28$ to $164$ in Eq.~\ref{eq:lorenz}, initially desynchronization, then intermittent complete synchronization, and finally complete synchronization states have been ascertained. Along with that, the ranges of oscillations are increasing with the increase in $r$-value. 
\begin{figure}[htbp!]
	\hspace*{10 mm}
	\includegraphics[width=35cm,height=8.2cm, keepaspectratio]{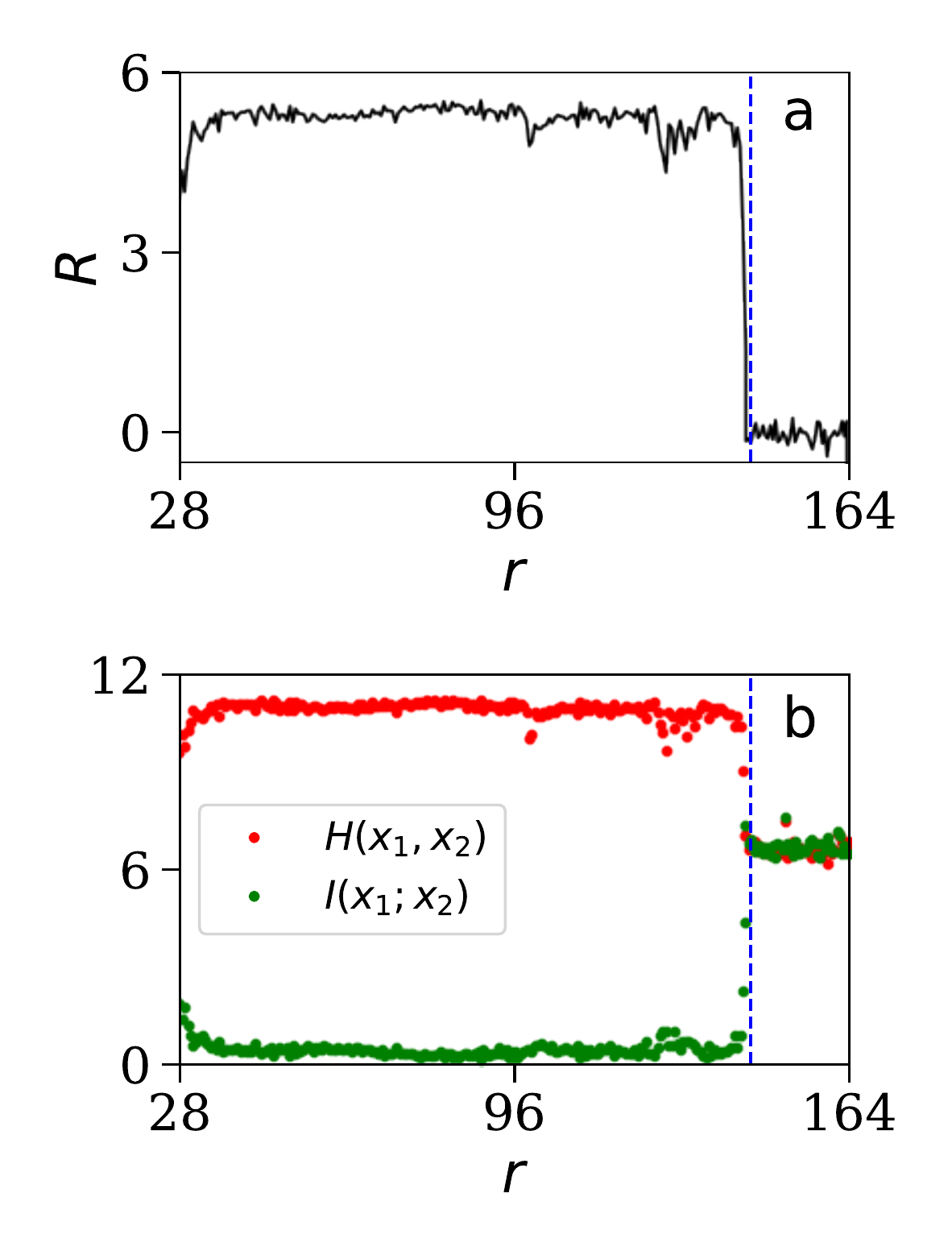}
	\caption{ \emph{(color online)} \textbf{Early detection of complete synchrony:} (a) The early detection index ($R$) is plotted as a function of $r$. $R$ reaches zero before the coupled oscillators yield the synchronized state. (b) The joint entropy and mutual information are plotted as a function of $r$. The vertical blue dashed-lines in both subplots are at $r = 144$. }
	\label{fig:cs_ew}
\end{figure}
Now, we are interested in calculating the early prediction index ($R$) of the coupled Lorenz oscillators (Eq.~\ref{eq:lorenz}) to anticipate the complete synchronization state. In this regard, we adopt the $x$-coordinates of Eq.~\ref{eq:lorenz} as the required variables to calculate $R$, i.e., $x_1$ and $x_2$ of Eq.~\ref{eq:lorenz} are the analogous variables of $X$ and $Y$ of Eq.~\ref{eq:order_para}. The variation of $R$ as a function of $r$ is plotted in Fig.~\ref{fig:cs_ew}a. It is observed that initially, $R$ has a higher value, and as $r$ increases $R$ decreases and becomes zero before the coupled oscillators yield the complete synchronized state. Being more explicit, $R$ reaches zero around $r = 144$ (the vertical blue dashed-line of Fig.~\ref{fig:cs_ew}a), and the coupled Lorenz oscillators yield the complete synchrony around $r = 164$ as shown in Fig.~\ref{fig:lorenz_cs}a. Also, we have plotted the joint entropy $H(x_1, x_2)$ (red plot) and the mutual information $I(x_1; x_2)$ (green plot) as a function of $r$ in Fig.~\ref{fig:cs_ew}b. We observe that, initially, during the desynchronized state, $H(x_1, x_2)$ and $I(x_1; x_2)$ are separated apart, reaching each other and overlapping as $r$ increases. Thus, $R$ decreases and saturates to zero before the synchrony, and its saturation to zero is an early indicator of complete synchrony. 
Thus, we have studied the transition to complete synchrony using the mutually coupled Lorenz oscillators and verified the applicability of the early detection index ($R$). Next, we extend our study to the transition to generalized synchrony in the coupled oscillator model.  
\subsubsection{Mutually coupled dynamo and Lorenz}
\label{sec:gs}
In this section, we switch to the second example of coupled systems: mutually coupled dynamo and Lorenz oscillators. Similar to the previous example, we increase one system parameter and study the change in dynamics of the interacting oscillators. The corresponding equations of motion are given by:
\begin{subequations}
	\label{eq:gs}
	\begin{eqnarray}
		\frac{d{x}_1}{dt} &=& y_1 z_1 -1.7 x_1,\label{eq:gs_a}\\
		\frac{d{y}_1}{dt} &=& (z_1-0.5)x_1 - 1.7 y_1,\label{eq:gs_b}\\
		\frac{d{z}_1}{dt} &=& 1-x_1 y_1 +\alpha (z_2 - z_1), \label{eq:gs_c}\\
		\frac{d{x}_2}{dt} &=& 10(y_2-x_2),\label{eq:gs_d}\\
		\frac{d{y}_2}{dt} &=& x_2(r-z_2)-y_2,\label{eq:gs_e}\\
		\frac{d{z}_2}{dt} &=& x_2 y_2 - \frac{8}{3}z_2 + \alpha (z_1 - z_2).\label{eq:gs_f}
	\end{eqnarray}
\end{subequations}
Following Eqs.~\ref{eq:1} and \ref{eq:gs}, $\mathbf{x}_1 (t)$ and $\mathbf{x}_2(t)$ represent the phase space variables of the dynamo and the Lorenz oscillators, respectively. Also, we mention that the coupling matrix $\textsf{C} = \text{diag} (0, 0, 1)$, and the control parameter is $\mu_2 = r$. The initial condition for Eq.~\ref{eq:gs} is chosen as $(-0.01, 1.01, 1, -0.1, 0, 1)$.

\begin{figure}[htbp!]
	\hspace*{0.21 mm}
	\includegraphics[width=60cm,height=9.1cm, keepaspectratio]{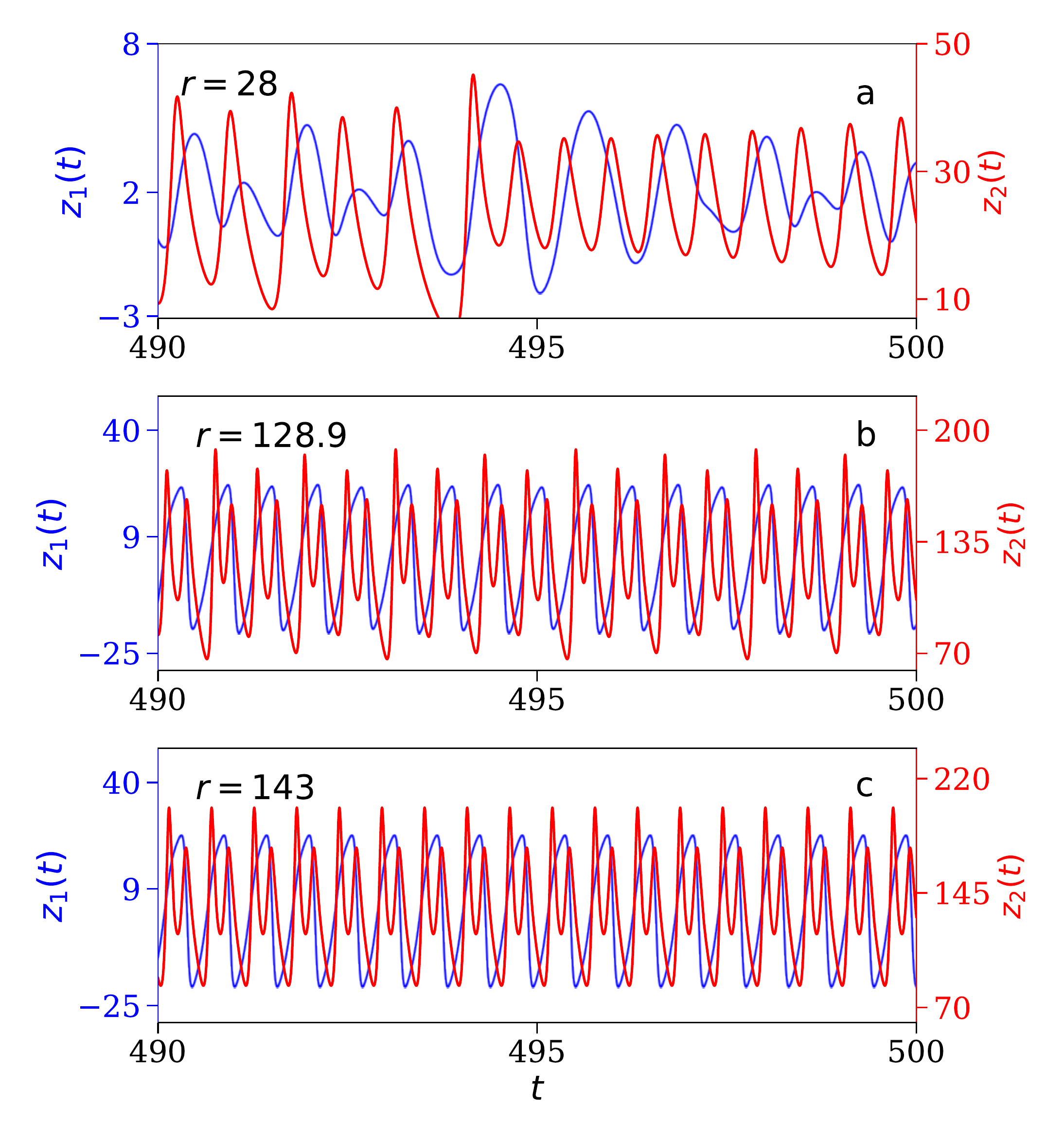}
	\caption{\emph{(color online)} \textbf{Aperiodic to periodic dynamics are detected for bidirectionally coupled dynamo and Lorenz oscillators:} The $z$-coordinates of Eq.~\ref{eq:gs} are plotted for three different values of $r$: $28$, $128.9$, and $143$. In subplot~(a), chaotic dynamics is observed in both the oscillators. In subplot~(b), both the time series becoming periodic intermittently, as the peaks have occasionally identical values. Finally, in subplot~(c), periodic dynamics is detected both in $z_1(t)$ and $z_2(t)$.}
	\label{fig:gs}
\end{figure}
Here also, we keep the coupling strength parameter fixed at $\alpha = 0.65$. Besides, we vary the system parameter $r$ (Eq.~\ref{eq:gs_e}) monotonically from $28$ to $143$ to investigate the corresponding changes in the dynamics. As the control parameter increases monotonically, the transition from chaotic to periodic dynamics for the interacting oscillators is detected. Both the oscillators are showing the chaotic dynamics and together lead to the desynchronized state at $r = 28$ (Fig.~\ref{fig:gs}a). As the control parameter $r$ increases further, at $r = 128.9$, the participating oscillators become periodic intermittently, as shown in Fig.~\ref{fig:gs}b. The magnitudes of peaks of $z_1(t)$ (or $z_2(t)$) are matching occasionally at $r = 128.9$. Hence, we call this dynamical state as the intermittent periodic state. Finally, in Fig.~\ref{fig:gs}c, corresponding to $r = 143$, both the oscillators become periodic in time $t$. Note that, at $r =143$, $z_1(t)$ and $z_2(t)$ have periodic oscillations in time $t$ of periods one and two respectively. Besides, the inter-peak separations of $z_1(t)$ and $z_2(t)$ become constant. Since $z_1(t)$ and $z_2(t)$ become periodic in time with the constant amplitudes, we can always write $z_1(t)$ as a function of $z_2(t)$, which further helps in writing $\mathbf{x_1} = \phi (\mathbf{x_2})$, where $\phi$ is a functional that connects the phase space variables of one oscillator with the other. In passing, note that using the concept of recurrence plots~\cite{romano05}, neither phase synchronization nor generalize synchronization is detected at $r =143$ (or at $r =128.9$). May be this tool of recurrence plots are applicable to observe synchronization in coupled identical oscillators with parameter mismatch~\cite{romano04,romano05,ghosh20_2}.
\begin{figure}[htbp!]
	\hspace*{10 mm}
	\includegraphics[width=35cm,height=8.2cm, keepaspectratio]{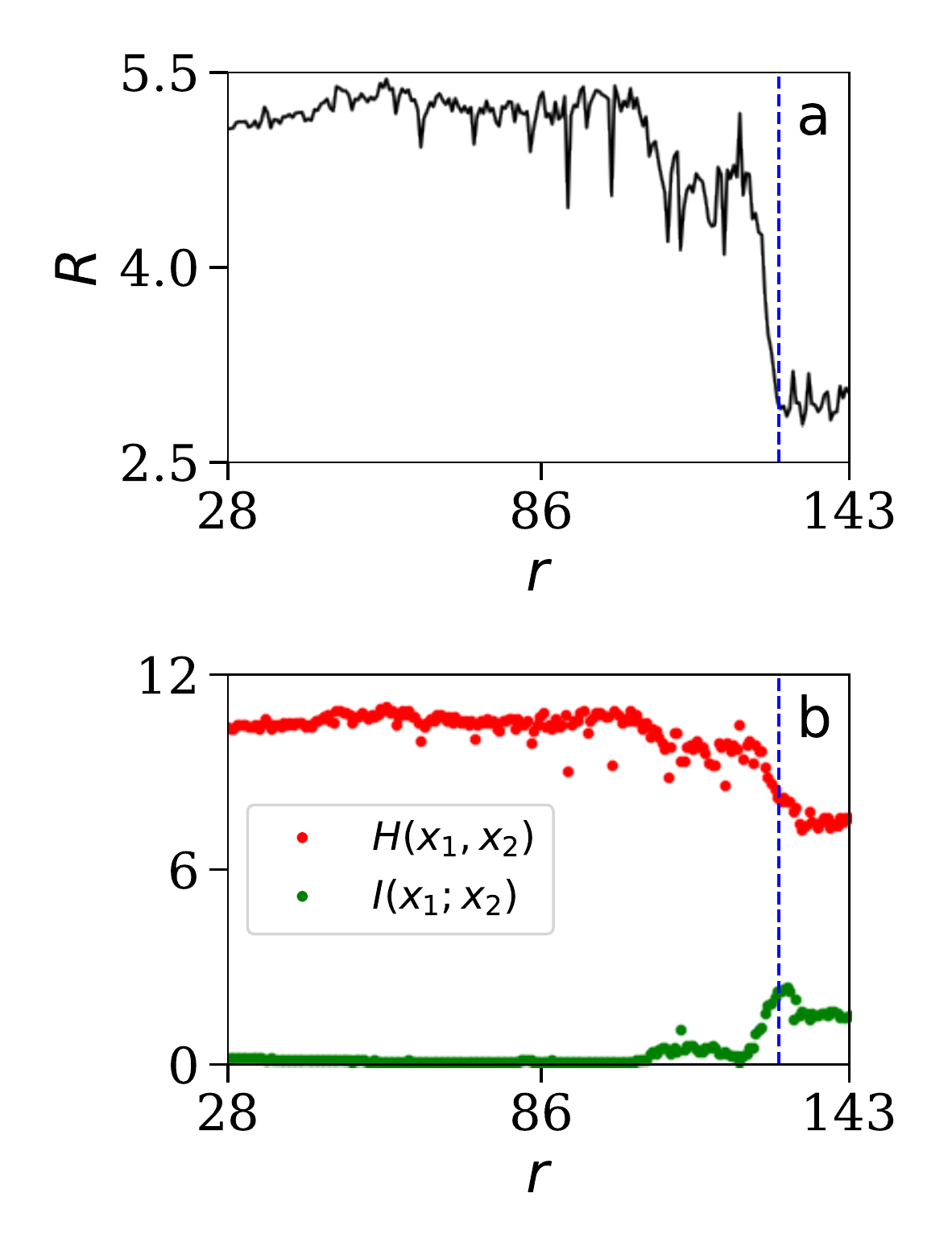}
	\caption{ \emph{(color online)} \textbf{Early detection of generalized synchrony:} (a) The early detection index ($R$) is plotted as a function of $r$. $R$ reaches a non-zero value and saturates before the coupled oscillator (Eq.~\ref{eq:gs}) yield the generalized synchronization state. (b) The joint entropy and mutual information are plotted as a function of $r$. The vertical blue dashed-lines in both subplots are drawn at $r = 130$.}
	\label{fig:gs_ew}
\end{figure}
Going further, we calculate the early detection index $R$ for coupled dynamo and Lorenz oscillators (Eq.~\ref{eq:gs}). Figure~\ref{fig:gs_ew}a depicts the variation of $R$ as a function of $r$. It is observed that initially, $R$ has a larger value, and as $r$ increases $R$ reduces to a small non-zero number and saturates before the coupled oscillators (Eq.~\ref{eq:gs}) yield the generalized synchrony. More explicitly, the aforementioned saturation starts at $r \simeq 130$, and the coupled oscillators attain the generalized synchrony at $r \simeq 143$ (Fig.~\ref{fig:gs}c). Besides, the separation between $H(x_1, x_2)$ (red plot) and $I(x_1; x_2)$ (green plot) decreases at larger values of $r$ and saturates before the generalized synchronized state in Eq.~\ref{eq:gs} as depicted in Fig.~\ref{fig:gs_ew}b. We note that, unlike Fig.~\ref{fig:cs_ew}a, $R$ saturates at a non-zero value for the generalized synchronization. It is because we are dealing with non-identical interacting oscillators for generalized synchrony. 
Until now, we have seen that during the transition to synchrony, the dynamics of individual oscillator switches from chaotic motion to periodic oscillation. We study this transition to synchrony using a different control parameter in Sec.~\ref{sec:alpha}. Following Eq.~\ref{eq:1}, the condition $\mu_1 = \mu_2$ is always satisfied from now onward. Each interacting oscillator exhibits chaotic dynamics during the synchronized state.
\subsection{Using the coupling strength as a control parameter}
\label{sec:alpha}

\begin{figure}[htbp!]
	\hspace*{2 mm}
	\includegraphics[width=40cm,height=8cm, keepaspectratio]{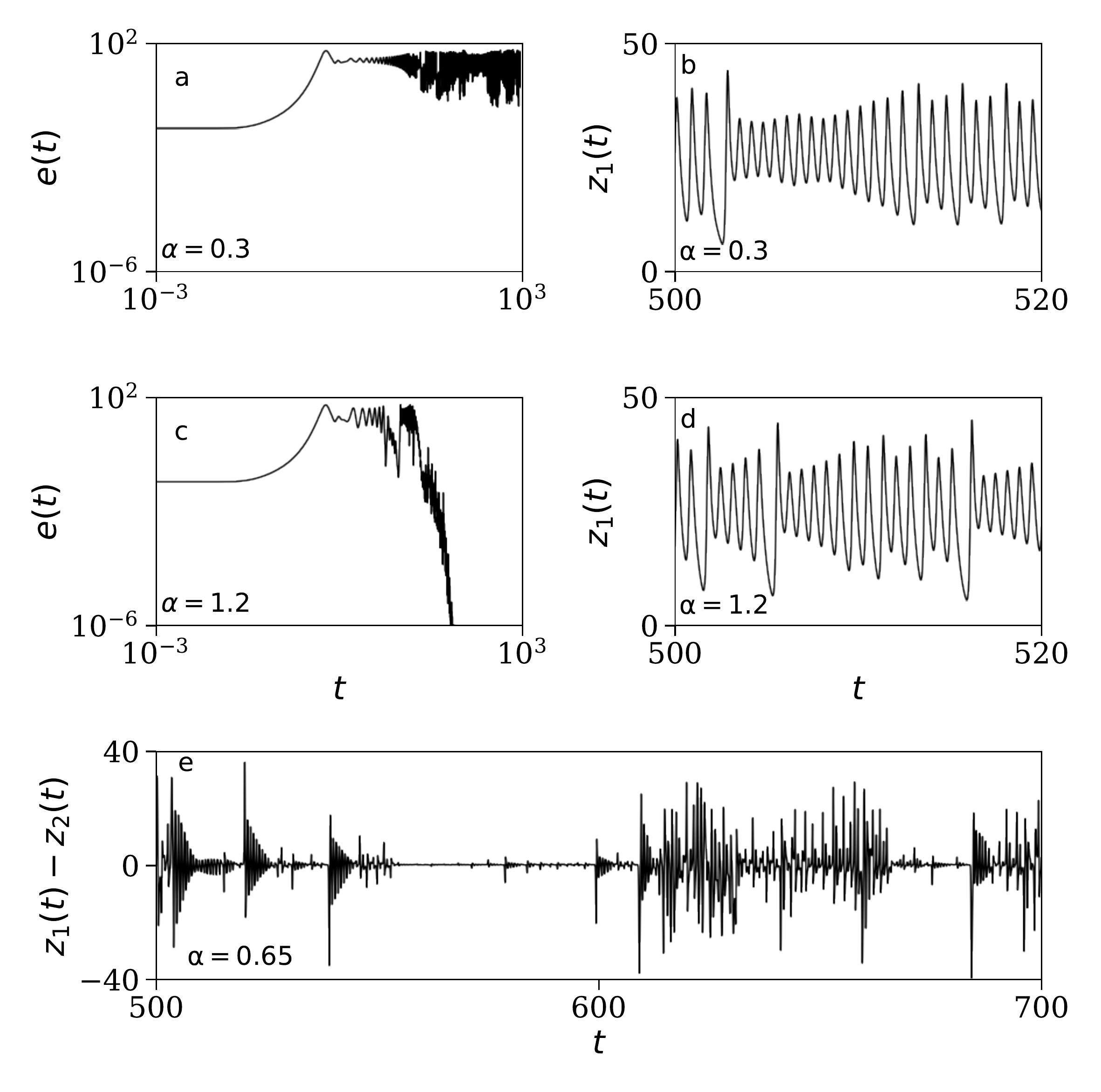}
	\caption{  \ag{ \textbf{Transition to complete synchrony using $\alpha$ as control parameter in coupled Lorenz oscillators model:} The variations of $e(t)$ (Eq.~\ref{eq:5}) in subplots (a) and (c) depict the desynchronization and complete synchronization states, respectively. Subplots (b) and (d) support the existence of chaotic dynamics at both values of $\alpha$. (e) The intermittent complete synchrony is detected at $\alpha = 0.65$.}}
	\label{fig:lor_alph_dyn}
\end{figure}

\ag{First, we consider the coupled Lorenz oscillators model (Eq.~\ref{eq:lorenz}). In this model, we keep the system parameter $r$ fixed at $28$ and vary the coupling strength $\alpha$ monotonically to study the transition to the complete synchronization. Also, we use the same initial condition that has already been mentioned in Sec.~\ref{sec:cs} to integrate Eq.~\ref{eq:lorenz}. In this regard, we increase $\alpha$ monotonically within the range $[0, 1.5]$. The complete synchronization is observed for $\alpha \geq 1.2$. However, the individual Lorenz oscillator exhibits chaotic dynamics over the entire range of $\alpha$ (i.e., $\alpha \in [0, 1.5]$). Figures~\ref{fig:lor_alph_dyn}a and \ref{fig:lor_alph_dyn}c depict the desynchronization and complete synchronization states, respectively, at two different values of $\alpha$. Besides, Figs.~\ref{fig:lor_alph_dyn}b and \ref{fig:lor_alph_dyn}d support that the first Lorenz oscillator exhibits chaotic dynamics at both values of $\alpha$. Similarly, we can plot $z_2(t)$ to confirm that the second Lorenz oscillator exhibits chaotic dynamics at different values of $\alpha$. Figures~\ref{fig:lor_alph_dyn}e shows the difference of the variables $z_1(t)$ and $z_2(t)$  becomes zero intermittently at $\alpha = 0.65$, which further implies the existence of the intermittent complete synchrony at $\alpha = 0.65$.}
\begin{figure}[htbp!]
	\hspace*{5 mm}
	\includegraphics[width=35cm,height=5.2cm, keepaspectratio]{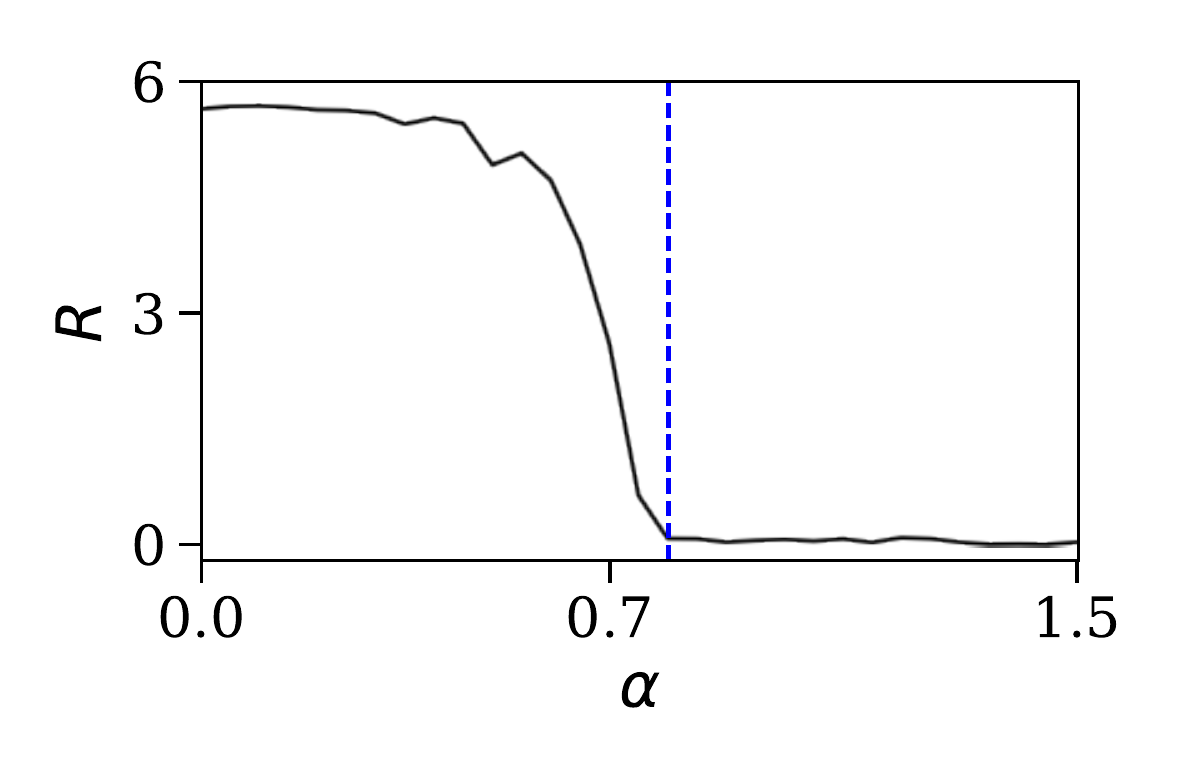}
	\caption{  \ag{ \emph{(color online)} \textbf{Early detection of complete synchrony using $\alpha$ as control parameter:} The early detection index ($R$) is plotted as a function of $\alpha$. $R$ reaches zero at $\alpha = 0.8$ and saturates before the coupled oscillators (Eq.~\ref{eq:lorenz}) yield the complete synchronization state. The vertical blue dashed-line is drawn at $\alpha = 0.8$.}}
	\label{fig:cs_ew_alpha}
\end{figure}
\ag{The early warning index $R$ of Eq.~\ref{eq:lorenz} has been calculated at different values of $\alpha$. Figure~\ref{fig:cs_ew_alpha} depicts the variation of calculated $R$ as a function of $\alpha$. $R$ reaches zero at $\alpha = 0.8$, whereas the transition to complete synchrony is detected at $\alpha = 1.2$. Therefore, based on Figs.~\ref{fig:cs_ew} and \ref{fig:cs_ew_alpha}, the concluding remark is that the early detection index is applicable for complete synchronization using both control parameters. }
\ag{Next, we are interested in studying the generalized synchrony in R\"ossler driven Lorenz oscillators, and the explicit form of equations of motion are given by:
\begin{subequations}
	\label{eq:gs_ross}
	\begin{eqnarray}
	\frac{d{x}_1}{dt} &=& -y_1 - z_1,\label{eq:gs_ross_a}\\
	\frac{d{y}_1}{dt} &=& x_1 + 0.2 y_1,\label{eq:gs_ross_b}\\
	\frac{d{z}_1}{dt} &=& 0.2 + z_1 (x_1 - 5.7),\label{eq:gs_ross_c}\\
	\frac{d{x}_2}{dt} &=& 16 (y_2 - x_2) + \alpha (x_1 - x_2),\label{eq:gs_ross_d}\\
	\frac{d{y}_2}{dt} &=& - x_2 z_2 + 45.92 x_2 - y_2,\label{eq:gs_ross_e}\\
	\frac{d{z}_2}{dt} &=& x_2 y_2 - 4 z_2.\label{eq:gs_ross_f}
	\end{eqnarray}
\end{subequations}
This model (Eq.~\ref{eq:gs_ross}) has already been used in literature to study the transition to generalized synchrony from desynchrony~\cite{abarbanel96}. In order to keep consistent with Eq.~\ref{eq:1}, $\mathbf{x_1}(t)$ and $\mathbf{x_2}(t)$ represents the phase space vectors of the R\"ossler and Lorenz oscillators, respectively. The coupling matrix $\textsf{C} = \text{diag}(1,0,0)$ and $\alpha$ in Eq.~\ref{eq:1a} is zero. Note that this R\"ossler driven Lorenz oscillators model is different from the previous three models because of the following two reasons: first, we deal with the unidirectional coupling, and second, $x$-coordinates (not the $z$-coordinates) of the interacting oscillators are coupled (Eq.~\ref{eq:gs_ross_d}).}

\begin{figure}[htbp!]
	\hspace*{5 mm}
	\includegraphics[width=35cm,height=5.2cm, keepaspectratio]{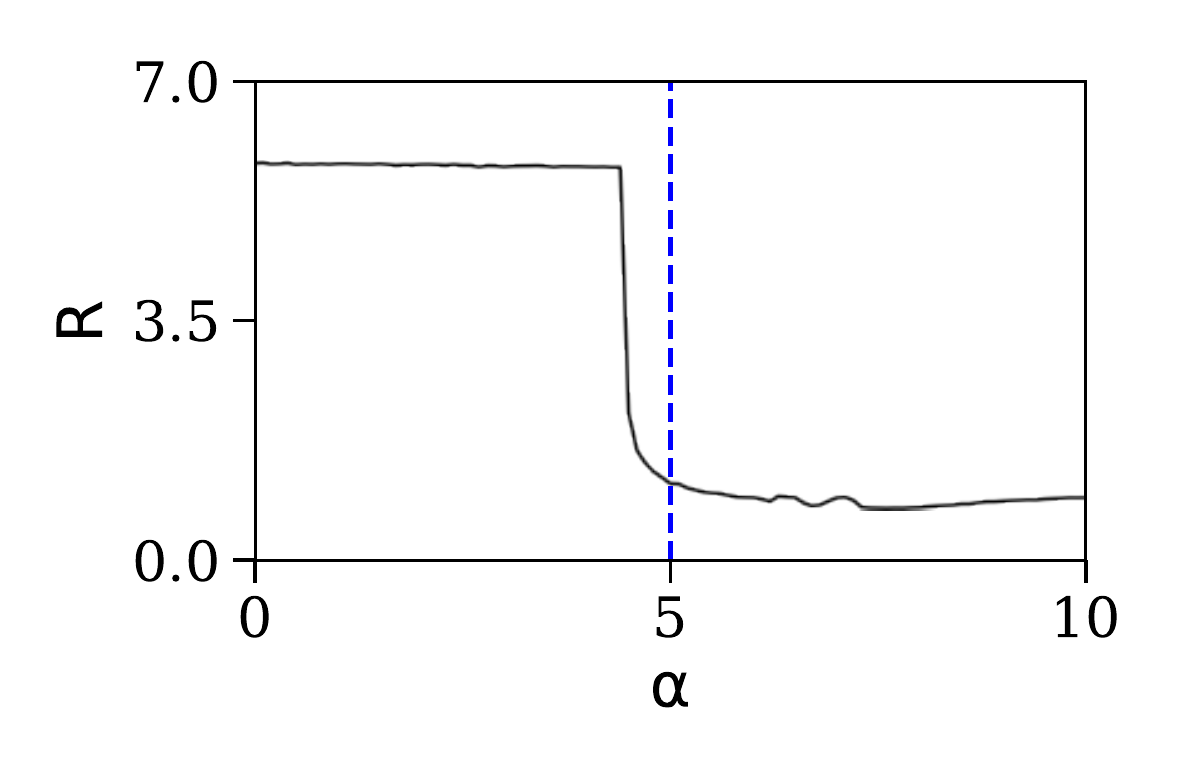}
	\caption{  \ag{ \emph{(color online)} \textbf{Early detection of generalized synchrony using $\alpha$ as control parameter:} The early detection index ($R$) is plotted as a function of the control parameter $\alpha$. $R$ reaches a non-zero number at $\alpha \simeq 5$ and almost saturates before the coupled oscillators (Eq.~\ref{eq:gs_ross}) yield the generalized synchronization state. The vertical blue dashed-line is drawn at $\alpha = 5$.}}
	\label{fig:gs_ew_alpha}
\end{figure}

\ag{The well-known `auxiliary system approach'~\cite{abarbanel96,pecora97,pikovsky01,balanov08} is generally used in the literature to study generalized synchrony of unidirectionally coupled oscillators. Following this method, we need to consider a second Lorenz oscillator, driven by the same R\"ossler oscillator, with the unaltered parameter values and has a different initial condition. This second Lorenz oscillator is also called the `auxiliary' driven oscillator. If $\mathbf{x'_2} (t) := (x'_2, y'_2, z'_2)$ represents the phase space vector of the auxiliary Lorenz oscillator, then the explicit form of the equations of motion are given by:
\begin{subequations}
	\label{eq:auxi}
	\begin{eqnarray}
	\frac{d{x}'_2}{dt} &=& 16 (y'_2 - x'_2) + \alpha (x_1 - x'_2),\label{eq:5a}\\
	\frac{d{y}'_2}{dt} &=& - x'_2 z'_2 + 45.92 x'_2 - y'_2,\label{eq:5b}\\
	\frac{d{z}'_2}{dt} &=& x'_2 y'_2 - 4 z'_2.\label{eq:5c}
	\end{eqnarray}
\end{subequations} 
The complete synchronization between $\mathbf{x_2} (t)$ and $\mathbf{x'_2} (t)$ confirms the generalized synchrony between oscillators $\mathbf{x_1} (t)$ and $\mathbf{x_2} (t)$. Similar to Eq.~\ref{eq:5}, we define the Euclidean norm between $\mathbf{x_2} (t)$ and $\mathbf{x'_2} (t)$ as follows:
\begin{equation}
	\label{eq:5_gs}
	e'(t):=||\mathbf{x}_2(t) - \mathbf{x}'_2(t)||.
\end{equation}
}
\begin{figure}[htbp!]
	\hspace*{0 mm}
	\includegraphics[width=35cm,height=8.5cm, keepaspectratio]{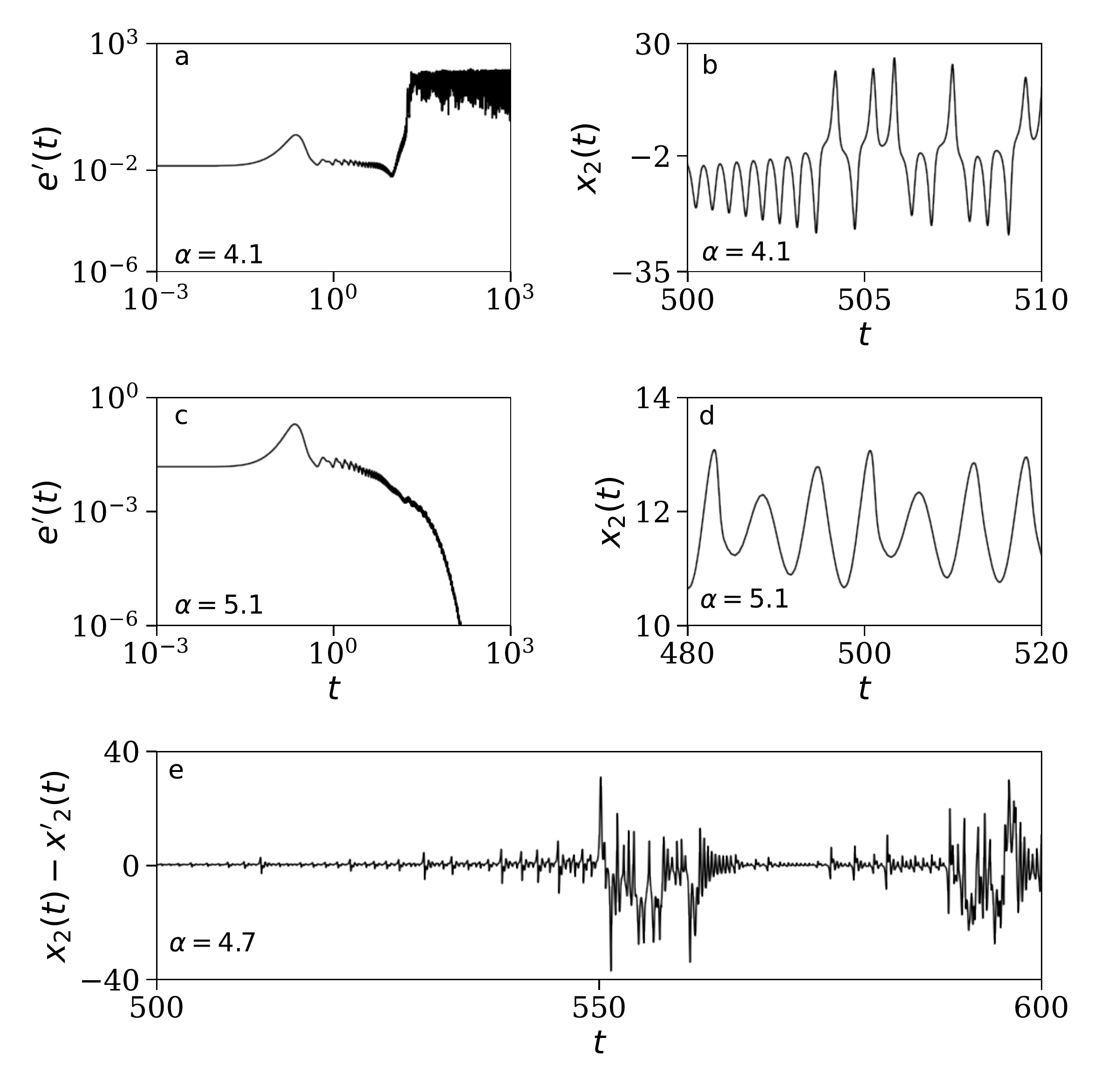}
	\caption{  \ag{ \textbf{Transition to complete synchrony between the Lorenz and the auxiliary Lorenz oscillators using $\alpha$ as control parameter:} The variations of $e'(t)$ (Eq.~\ref{eq:5_gs}) in subplots (a) and (c) depict the desynchronization and complete synchronization states, respectively. Subplots (b) and (d) support the existence of chaotic dynamics during the desynchronized and complete synchronized states. (e) The intermittent complete synchrony between $\mathbf{x_2} (t)$ and $\mathbf{x'_2} (t)$ is detected at $\alpha = 4.7$.}}
	\label{fig:gs_dynamics_alpha}
\end{figure}
\ag{In the numerical analysis of Eq.~\ref{eq:gs_ross}, we have adopted the initial condition as $(-9,0,0,0.01,-0.1, 0)$, and we choose $(0.01, -0.11, 0.01)$ as the initial condition of Eq.~\ref{eq:auxi}. Figures~\ref{fig:gs_dynamics_alpha}a and \ref{fig:gs_dynamics_alpha}c infer the desynchronized and the complete synchrony between $\mathbf{x_2} (t)$ and $\mathbf{x'_2} (t)$ at two different values of $\alpha$. These observations further imply the generalized synchrony between $\mathbf{x_2} (t)$ and $\mathbf{x_1} (t)$ at $\alpha = 5.1$. Also, the chaotic dynamics of the driven Lorenz oscillator at two different values of $\alpha$ are shown in Figs.~\ref{fig:gs_dynamics_alpha}b and \ref{fig:gs_dynamics_alpha}d. The intermittent complete synchronization between $\mathbf{x_2} (t)$ and $\mathbf{x'_2} (t)$ is detected at $\alpha = 4.7$ (Fig.~\ref{fig:gs_dynamics_alpha}e). The early detection index ($R$) of Eq.~\ref{eq:gs_ross} has been calculated as a function of $\alpha$. The variation of $R$ as a function of $\alpha$ is shown in Fig.~\ref{fig:gs_ew_alpha}. $R$ reaches a non-zero small number as we increase $\alpha$ monotonically and gets almost saturated for $\alpha \geq 5$. Therefore, for this example, we have anticipated the generalized synchrony using the coupling strength as a control parameter.}
\ag{At the end of this section, we have seen that the early detection index is suitable to detect complete and generalized synchronizations early using both control parameters. To this end, we mention that we have studied the applicability of $R$ in coupled Chen~\cite{chen99} oscillators in Appendix~\ref{sec:chen}. Also, the robustness of $R$ has been investigated using a coupled R\"ossler oscillators model in the presence of noise (Appendix~\ref{sec:ross}). Finally, we extend our study to a network of coupled R\"ossler oscillators in Appendix~\ref{sec:ring}.}
\section{Conclusion and discussions}
\label{sec:conclusion}
\ag{This paper has focused on the applicability of an \emph{early warning index} while studying the transitions from desynchrony to synchrony in various coupled oscillator models using two different control parameters. Initially, we have chosen an \emph{unconventional} system parameter as the required control parameter. Consequently, in the first example, we have investigated the transition to complete synchrony using the mutually coupled Lorenz oscillators. We initially obtained the desynchronized state between the coupled oscillators, and both the individual oscillators show chaotic dynamics. Further increase in the bifurcation parameter leads to intermittent complete synchronization between the oscillators, followed by complete synchronization. It is observed that the early detection index reached zero and saturated before the interacting oscillators yielded the complete synchronized state. On the same track, for the example of a mutually coupled dynamo and Lorenz oscillators, we have first ascertained chaotic desynchronization, then generalized synchronization through the route of intermittency. Similar to the example of coupled Lorenz oscillators, the transition to periodic dynamics from chaos is also ascertained for mutually coupled dynamo and Lorenz oscillators. This transition to generalized synchrony is also anticipated using the early detection index.}   
\ag{In the next part, we have chosen the coupling strength as the required control parameter for the transition to synchrony from desynchrony. The applicability of the early detection index is verified. We have chosen the examples of coupled Lorenz oscillators and R\"ossler driven Lorenz oscillators to study the transitions to complete and generalized synchronizations, respectively. Unlike the previous case, the individual oscillator exhibits chaotic dynamics during the synchronized state using this control parameter. Furthermore, the robustness of the early warning index has been established using an example of coupled R\"ossler oscillators in the presence of noise. Finally, we have extended our study to a network of ten interacting R\"ossler oscillators to study the transition to complete synchrony and verified the applicability of the early warning index.}
%

%
%

%
\ag{Studying synchronizations in coupled oscillator models varying the coupling strength parameter is popular in the literature. On the contrary, use of a different control parameter (other than the coupling strength) is relevant in experiments, such as thermoacoustic~\cite{sujith21}, aeroacoustic~\cite{mondal17_chaos}, and aeroelastic~\cite{raaj19} systems. The early warning index used in this paper could be suitable for these experiments as this index is convenient for experimental data and mathematical models. In addition, epilepsy and Parkinson's diseases~\cite{dominguez05,hammond07,rubchinsky12} are mainly because of the synchronized firing of neuron oscillators. The extension of the applicability of this early warning index in studying neural dynamics is also an exciting direction for future work.}
\section*{Acknowledgment}
The author thanks Prof. R. I. Sujith and Dr. S. Sur for several fruitful discussions and the anonymous referees for their constructive criticisms. The author gratefully acknowledges the Institute Post-Doctoral Fellowship of the Indian Institute of Technology Madras, India.

\section*{Data Availability Statements}
The data that support the findings of this study are available from the corresponding author upon reasonable request.

\appendix
\section{Coupled Chen oscillators}
\label{sec:chen}
\ag{In this section, we consider the example of mutually coupled Chen~\cite{chen99} oscillators. The explicit form of the equations of motion is given by:  
\begin{subequations}
	\label{eq:chen}
	\begin{eqnarray}
	\frac{d{x}_1}{dt} &=& a(y_1 - x_1),\label{eq:chen_a}\\
	\frac{d{y}_1}{dt} &=& (c-a)x_1 -x_1 z_1 + cy_1 + \alpha (y_2-y_1),\label{eq:chen_b}\\
	\frac{d{z}_1}{dt} &=& y_1 x_1 - bz_1,\label{eq:chen_c}\\
	\frac{d{x}_2}{dt} &=& a(y_2 - x_2),\label{eq:chen_d}\\
	\frac{d{y}_2}{dt} &=& (c-a)x_2 -x_2 z_2 + cy_2 + \alpha (y_1-y_2),\label{eq:chen_e}\\
	\frac{d{z}_2}{dt} &=& y_2 x_2 - bz_2,\label{eq:chen_f}
	\end{eqnarray}
\end{subequations}
\begin{figure}[htbp!]
	\hspace*{5 mm}
	\includegraphics[width=35cm,height=9.2cm, keepaspectratio]{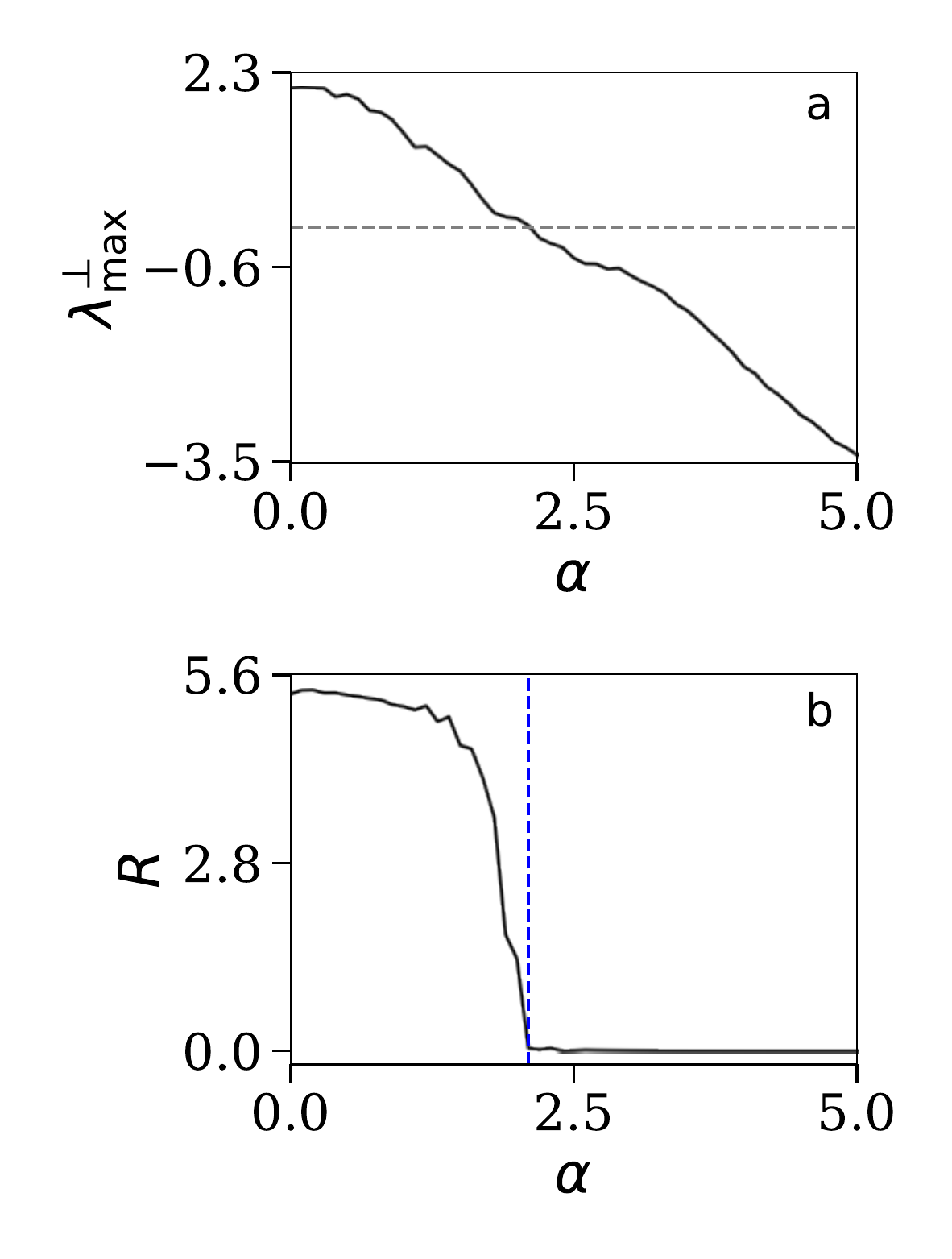}
	\caption{ \ag{ \emph{(color online)} \textbf{Early detection of complete synchrony in coupled Chen oscillators:} (a) The maximum conditional Lyapunov exponent ($\lambda^{\bot}_{\rm max}$) is plotted as a function of the coupling strength ($\alpha$). $\lambda^{\bot}_{\rm max}$ becomes negative at $\alpha = 2.14$. The horizontal gray dashed-line corresponds to $\lambda^{\bot}_{\rm max} = 0$. (b) The early detection index ($R$) is plotted as a function of the control parameter $\alpha$. $R$ reaches zero at $\alpha \simeq 2.1$ and saturates before the coupled oscillators (Eq.~\ref{eq:chen}) yield the generalized synchronization state. The vertical blue dashed-line is drawn at $\alpha = 2.1$.}}
	\label{fig:cs_ew_chen}
\end{figure}
with $a=35$, $b = 3$, and $c = 28$. Here, we have coupled the $y$-coordinates of the interacting oscillators. In order to study the transition to complete synchrony, the coupling strength is chosen as a control parameter. Also, we have calculated the maximum conditional Lyapunov exponent ($\lambda^{\bot}_{\rm max}$)~\cite{pecora97} of Eq.~\ref{eq:chen} for different values of $\alpha$. By definition, the negativity of $\lambda^{\bot}_{\rm max}$ implies the complete synchronization state. The maximum conditional Lyapunov exponent is a suitable measure to detect the complete synchronization between two diffusively coupled identical oscillators, where the explicit form of equations of motion is available.} 
\ag{In analysis, the initial condition to integrate Eq.~\ref{eq:chen} is chosen as $(-10, 0, 37,-10.1, 0.1, 37.1)$. The maximum conditional Lyapunov exponent becomes negative for $\alpha \geq 2.14$ (Fig.~\ref{fig:cs_ew_chen}a). Hence, the coupled Chen oscillators yield the complete synchronized state at $\alpha = 2.14$. Besides, Fig.~\ref{fig:cs_ew_chen}b depicts that $R$ reaches zero at $\alpha= 2.10$ and saturates. We use the coordinates $x_1(t)$ and $x_2(t)$ to calculate $R$ at each $\alpha$. }
\section{Coupled R\"ossler oscillators with noise}
\label{sec:ross}
\begin{figure}[htbp!]
	\hspace*{3 mm}
	\includegraphics[width=35cm,height=8.2cm, keepaspectratio]{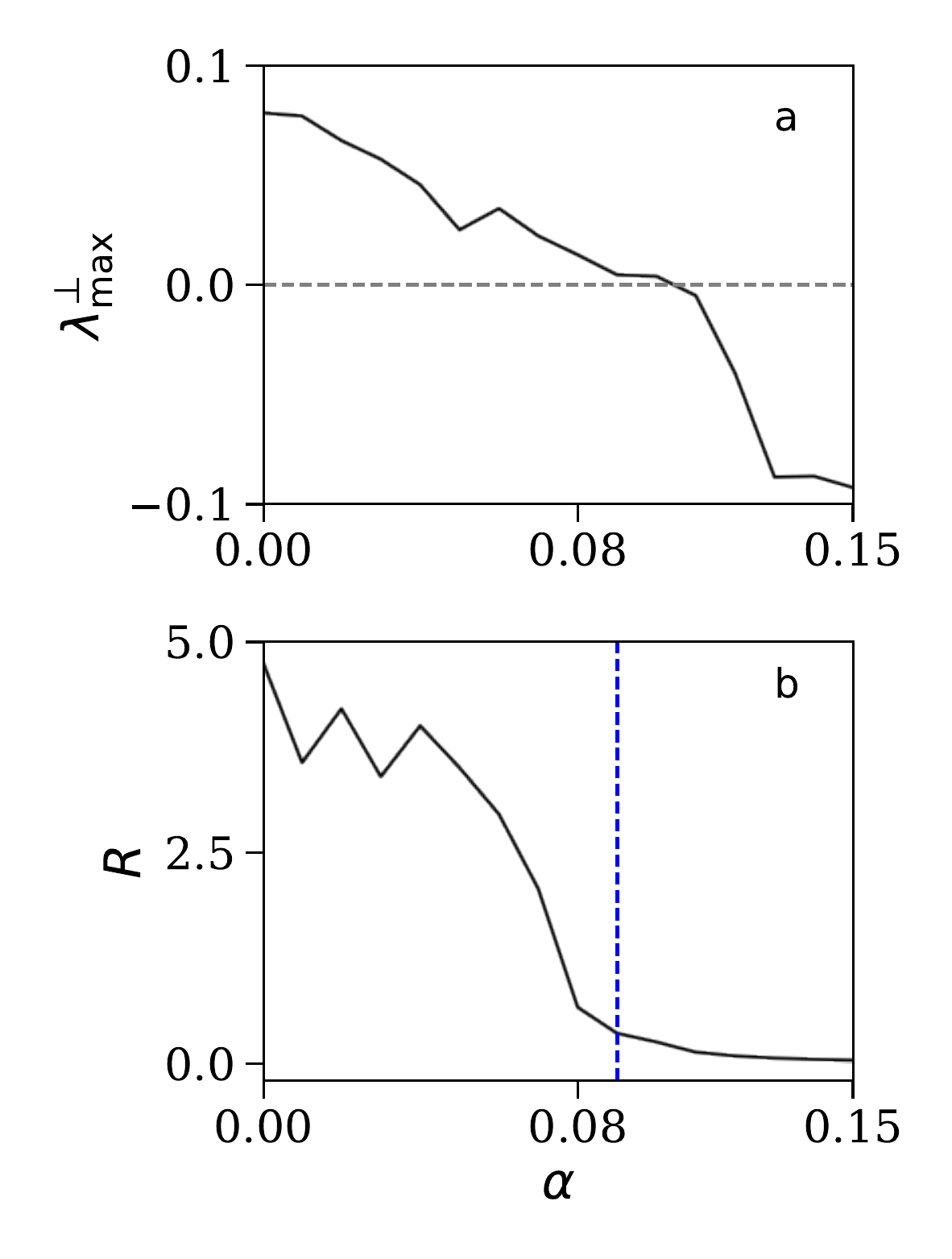}
	\caption{  \ag{ \emph{(color online)} \textbf{Robustness of the early detection index in the presence of noise:} (a) The maximum conditional Lyapunov exponent ($\lambda^{\bot}_{\rm max}$) becomes negative for $\alpha \geq 1.1$. The horizontal gray dashed-lines corresponds to $\lambda^{\bot}_{\rm max} = 0$. (b) The early detection index ($R$) is plotted as a function of the control parameter $\alpha$ in the presence of noise with noise strength $D = 1 \times 10^{-5}$. $R$ reaches a non-zero small number and saturates before the coupled oscillators (Eq.~\ref{eq:ross_ring}) yield the complete synchronization state. The vertical blue dashed-line is drawn at $\alpha = 0.09$.}}
	\label{fig:ew_ross}
\end{figure}
\ag{Now, the example of two coupled R\"ossler oscillators is considered in the presence of noise. The explicit form of the coupled oscillators are as follow:
\begin{subequations}
	\label{eq:ross}
	\begin{eqnarray}
	\frac{d{x}_{1,2}}{dt} &=& -y_{1,2} - z_{1,2} + \alpha (x_{2,1} - x_{1,2}) + D \eta(t),\label{eq:ross_a}\\
	\frac{d{y}_{1,2}}{dt} &=& x_{1,2} + 0.2 y_{1,2},\label{eq:ross_b}\\
	\frac{d{z}_{1,2}}{dt} &=& 0.2 + z_{1,2} (x_{1,2} - 5.7).\label{eq:ross_c}
	\end{eqnarray}
\end{subequations} 
We introduce the noise term in Eq.~\ref{eq:ross_a}. Here, $D$ represents the noise amplitude and $\eta(t)$ is adopted from a Gaussian distribution (of zero mean and unit variance) randomly, i.e., $ \left\langle \eta(t) \eta(t') \right\rangle = \delta(t-t')$.
}
\ag{The initial condition for Eq.~\ref{eq:ross} is chosen as $(-9, 0, 0,\\ -9.1, 0.1, 0)$. The maximum conditional Lyapunov exponent ($\lambda^{\bot}_{\rm max}$) of Eq.~\ref{eq:ross} with $D = 0$ becomes negative at $\alpha \simeq 0.11$ (Fig.~\ref{fig:ew_ross}a). It implies that coupled R\"ossler oscillators (Eq.~\ref{eq:ross}) lead the complete synchronized state for $\alpha \geq 0.11$ in the absence of noise (i.e., $D = 0$). Figure~\ref{fig:ew_ross}b depicts the variations of $R$ as a function of $\alpha$ at $D = 1 \times 10^{-5}$. For a fixed value of $\alpha$, the plotted $R$ in Fig.~\ref{fig:ew_ross}b is the average of all $R$ values calculated over $50$ realizations of the noise $\eta(t)$. $R$ reaches a small number at $\alpha \simeq 0.09$ and  almost saturates. Therefore, Fig.~\ref{fig:ew_ross} helps to conclude that the applicability of $R$ is robust in the presence of noise.}  
\section{Network of R\"ossler oscillators}
\label{sec:ring}
\begin{figure}[htbp!]
	\hspace*{0 mm}
	\includegraphics[width=35cm,height=5.5cm, keepaspectratio]{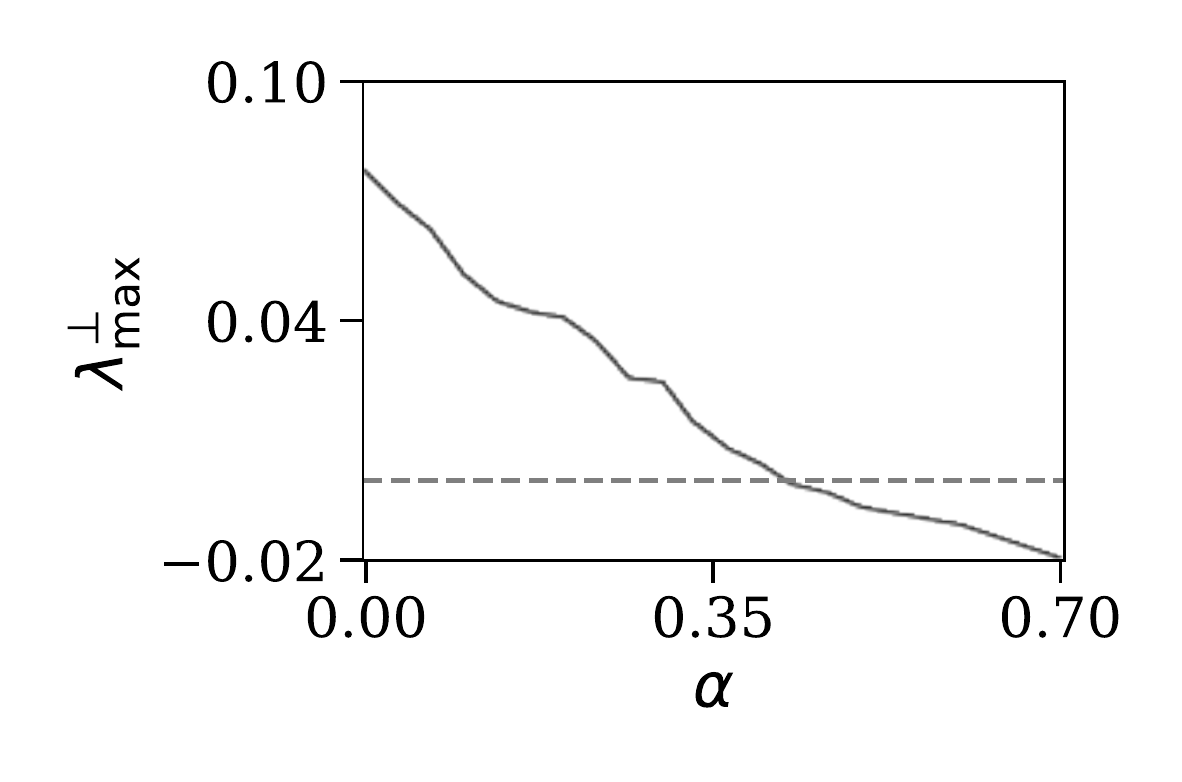}
	\caption{  \ag{ \emph{(color online)} \textbf{Transition to complete synchrony in a ring of coupled R\"ossler oscillators:} The maximum conditional Lyapunov exponent ($\lambda^{\bot}_{\rm max}$) of Eq.~\ref{eq:ross_ring} is plotted as a function of the control parameter $\alpha$, and $\lambda^{\bot}_{\rm max}$ becomes negative at $\alpha = 0.41$. It implies that the coupled oscillators (Eq.~\ref{eq:ross_ring}) yield the complete synchronization state for $\alpha \geq 0.41$. The horizontal gray dashed-line is drawn at $\lambda^{\bot}_{\rm max} = 0$.}}
	\label{fig:ross_ring_cle}
\end{figure}
\begin{figure}[htbp!]
	\hspace*{3 mm}
	\includegraphics[width=35cm,height=13.2cm, keepaspectratio]{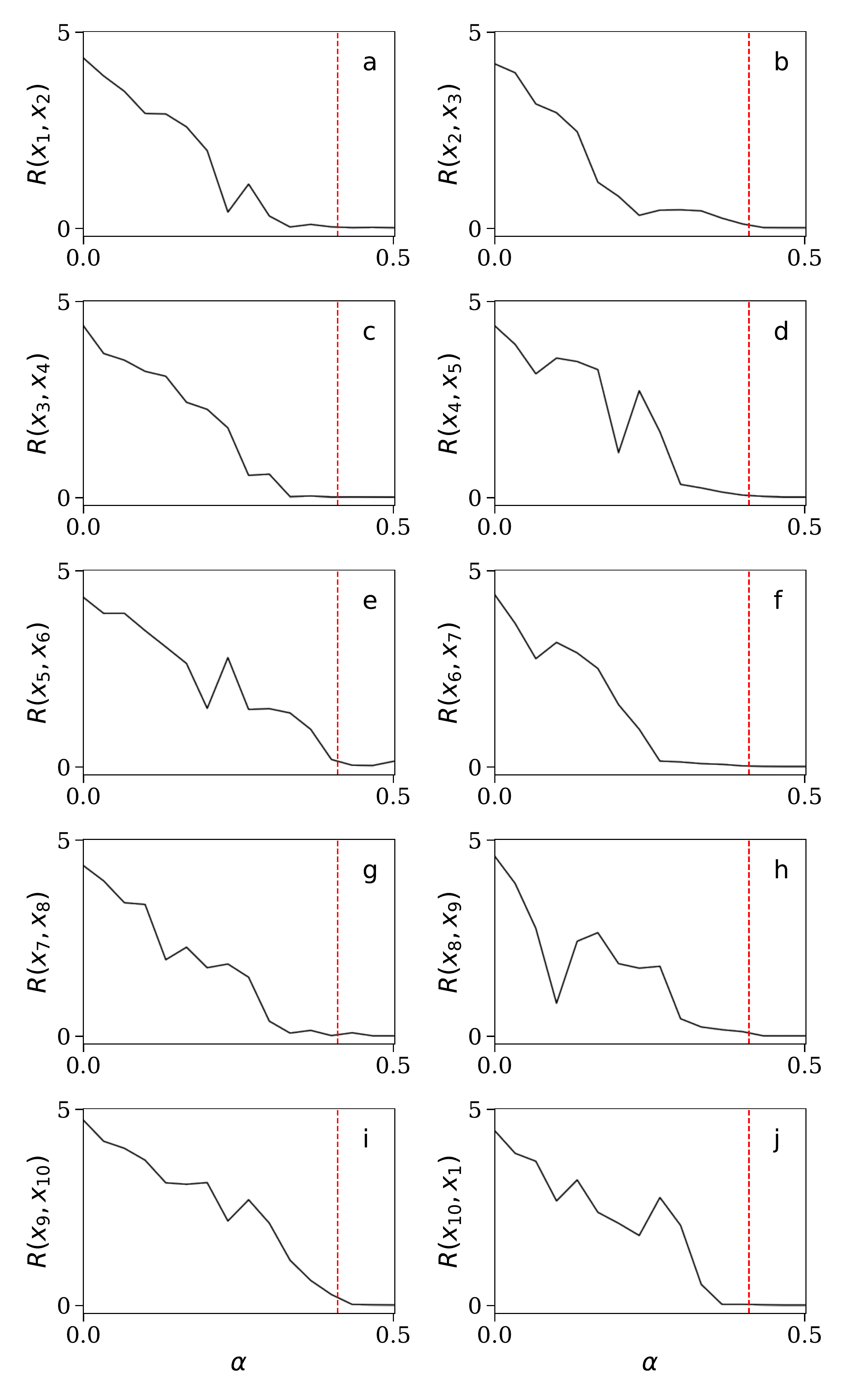}
	\caption{  \ag{ \emph{(color online)} \textbf{Early detection of complete synchrony in a ring of coupled R\"ossler oscillators:} The early detection index ($R$) is plotted as a function of the control parameter $\alpha$ using the $x$-coordinates of the interacting oscillators. In most cases, $R$ reaches zero and saturates before the coupled oscillators (Eq.~\ref{eq:ross_ring}) yield the complete synchronization state. The vertical red dashed-lines in all subplots are at $\alpha = 0.41$.}}
	\label{fig:ew_ross_ring}
\end{figure}

\ag{Until now, we have studied different examples of two coupled oscillators. This section extends our study to a network of interacting oscillators. We consider a ring of $N$ mutually coupled R\"ossler oscillators. Following is the equations of motion of the $i^{\rm th}$ oscillator:
\begin{subequations}
	\label{eq:ross_ring}
	\begin{eqnarray}
	\frac{d{x}_i}{dt} &=& -y_i - z_i + \alpha (x_{i+1} + x_{i-1} - 2x_i),\label{eq:3a}\\
	\frac{d{y}_i}{dt} &=& x_i + 0.2 y_i,\label{eq:3b}\\
	\frac{d{z}_i}{dt} &=& 0.2 + z_i (x_i - 5.7),\label{eq:3c}
	\end{eqnarray}
\end{subequations} 
where $i = 1, 2, \cdots, N$, with $x_{N+1} = x_1$ and $x_{-1} = x_N$. Here, the coupling strength $\alpha$ is the required control parameter, and we vary $\alpha$ monotonically within the range $[0, 0.5]$. In analysis, we adopt $N = 10$ and the initial conditions of interacting $10$ oscillators are chosen randomly from an uniform distribution with the boundaries $-0.1$ and $0.1$. The variation of $\lambda^{\bot}_{\rm max}$ as a function of $\alpha$ infers that all $10$ oscillators yield complete synchrony for $\alpha \geq 0.41$ (Fig.~\ref{fig:ross_ring_cle}). Besides, in Fig.~\ref{fig:ew_ross_ring}, we have plotted the early detection index $R$ using $x$-coordinates of the neighbour oscillators. An overall decreasing nature of $R$ is detected in all cases and $R$ saturates at zero in most cases. To this end, note that over this range of $\alpha$ (i.e., $\alpha \in [0, 0.5]$), each interacting R\"ossler oscillator exhibits chaotic oscillation.}    
\bibliographystyle{unsrt}
\bibliography{Ghosh_manuscript.bib}

\begin{thebibliography}{10}

\bibitem{winfree01}
A.~T. Winfree.
\newblock {\em The {G}eometry of {B}iological {T}ime}.
\newblock Springer Press, New York, first edition, 2001.

\bibitem{lakshmanan03}
M.~Lakshmanan and S.~Rajasekar.
\newblock {\em Nonlinear Dynamics: Integrability, Chaos, and Patterns}.
\newblock Springer Press, New York, first edition, 2003.

\bibitem{balanov08}
A.~Balanov, N.~Janson, D.~Postnov, and O.~Sosnovtseva.
\newblock {\em Synchronization: From Simple to Complex}.
\newblock Springer Press, Berlin, first edition, 2008.

\bibitem{feingold13}
G.~Feingold and I.~Koren.
\newblock A model of coupled oscillators applied to the
  aerosol–cloud–precipitation system.
\newblock {\em Nonlin. Processes Geophys.}, 20:1011, 2013.

\bibitem{muraki18}
Y.~Muraki.
\newblock {Application of a Coupled Harmonic Oscillator Model to Solar Activity
  and El Ni{\~n}o Phenomena}.
\newblock {\em J. Astron. Space Sci.}, 35:75, 2018.

\bibitem{go10}
C.~K.~C. Go and J.~T. Maquiling.
\newblock Using coupled harmonic oscillators to model some greenhouse gas
  molecules.
\newblock {\em AIP Conf. Proc.}, 1263:219, 2010.

\bibitem{miller17}
A.~J. Muraki~et al.
\newblock Coupled ocean–atmosphere modeling and predictions.
\newblock {\em J. Mar. Res.}, 75:361, 2017.

\bibitem{pikovsky01}
A.~Pikovsky, M.~Rosenblum, and J.~Kurths.
\newblock {\em Synchronization: A Universal Concept in Nonlinear Sciences}.
\newblock Cambridge University Press, New York, first edition, 2001.

\bibitem{bocc02}
S.~Boccaletti, J.~Kurths, G.~Osipov, D.~L. Valladares, and C.~S. Zhou.
\newblock The synchronization of chaotic systems.
\newblock {\em Phys. Rep.}, 366:1, 2002.

\bibitem{strogatz03}
S.~H. Strogatz.
\newblock {\em Sync: {T}he {E}merging {S}cience of {S}pontaneous {O}rder}.
\newblock Hyperion Press, New York, first edition, 2003.

\bibitem{ma05}
T.~Ma and S.~Wang.
\newblock {\em Bifurcation {T}heory and {A}pplications}.
\newblock World Scientific Press, Singapore, first edition, 2005.

\bibitem{strogatz07}
S.~H. Strogatz.
\newblock {\em Nonlinear Dynamics and Chaos: With Applications to Physics,
  Biology, Chemistry, and Engineering.}
\newblock CRC Press, India, second edition, 2014.

\bibitem{pc1990}
L.~M. Pecora and T.~L. Carroll.
\newblock Synchronization in chaotic systems.
\newblock {\em Phys. Rev. Lett.}, 64:821, 1990.

\bibitem{pecora97}
L.~M. Pecora, T.~L. Carroll, G.~A. Johnson, D.~J. Mar, and J.~F. Heagy.
\newblock Fundamentals of synchronization in chaotic systems, concepts, and
  applications.
\newblock {\em Chaos}, 7:520, 1997.

\bibitem{arenas08}
A.~Arenas, A.~Díaz-Guilera, J.~Kurths, Y.~Moreno, and C.~Zhou.
\newblock Synchronization in complex networks.
\newblock {\em Phys. Rep.}, 469:93, 2008.

\bibitem{eroglu17}
D.~Eroglu, J.~S.~W. Lamb, and T.~Pereira.
\newblock Synchronisation of chaos and its applications.
\newblock {\em Contemp. Phys.}, 58:207, 2017.

\bibitem{ghosh18}
A.~Ghosh, P.~Godara, and S.~Chakraborty.
\newblock Understanding transient uncoupling induced synchronization through
  modified dynamic coupling.
\newblock {\em Chaos}, 28:053112, 2018.

\bibitem{ghosh18_2}
A.~Ghosh, T.~Shah, and S.~Chakraborty.
\newblock Occasional uncoupling overcomes measure desynchronization.
\newblock {\em Chaos}, 28:123113, 2018.

\bibitem{sur20}
S.~Sur and A.~Ghosh.
\newblock Quantum counterpart of measure synchronization: {A} study on a pair
  of {H}arper systems.
\newblock {\em Phys. Lett. A}, 384:126176, 2020.

\bibitem{ghosh20}
A.~Ghosh and S.~Chakraborty.
\newblock Comprehending deterministic and stochastic occasional uncoupling
  synchronizations through each other.
\newblock {\em Eur. Phys. J. B}, 93:113, 2020.

\bibitem{dominguez05}
L.~G. Dominguez, R.~A. Wennberg, W.~Gaetz, D.~Cheyne, O.~C. Snead, and J.~L.~P.
  Velazquez.
\newblock Enhanced synchrony in epileptiform activity? local versus distant
  phase synchronization in generalized seizures.
\newblock {\em J. Neurosci.}, 25:8077, 2005.

\bibitem{hammond07}
C.~Hammond, H.~Bergman, and P.~Brown.
\newblock Pathological synchronization in parkinson's disease: networks, models
  and treatments.
\newblock {\em Trends Neurosci.}, 30:357, 2007.

\bibitem{rubchinsky12}
L.~L. Rubchinsky, C.~Park, and R.~M. Worth.
\newblock Intermittent neural synchronization in {P}arkinson’s disease.
\newblock {\em Nonlinear Dyn.}, 68:329, 2012.

\bibitem{lieuwen05}
T.~C. Lieuwen and V.~Yang.
\newblock {\em Combustion Instabilities in Gas Turbine Engines (Operational
  Experience, Fundamental Mechanisms and Modeling)}, volume 210.
\newblock Progress in Astronautics and Aeronautics, AIAA, 2005.

\bibitem{culick06}
F.~E.~C. Culick.
\newblock Unsteady motions in combustion chambers for propulsion systems.
\newblock Technical report, AGARDograph, NATO/RTO-AG-AVT-039, 2006.

\bibitem{fisher09}
S.C. Fisher, S.A. Rahman, and NASA~History Division.
\newblock {\em Remembering the Giants: {A}pollo Rocket Propulsion Development}.
\newblock Monographs in aerospace history. National Aeronautics and Space
  Administration, NASA History Division, Office of External Relations, 2009.

\bibitem{sujith21}
R.~I. Sujith and S.~A. Pawar.
\newblock {\em {T}hermoacoustic {I}nstability: {A} {C}omplex {S}ystems
  {P}erspective}.
\newblock Springer, Switzerland, 2021.

\bibitem{strogatz05}
S.~H. Strogatz, D.~M. Abrams, A.~McRobie, B.~Eckhardt, and E.~Ott.
\newblock Theoretical mechanics: Crowd synchrony on the {Millennium Bridge}.
\newblock {\em Nature}, 438:43, 2005.

\bibitem{ghosh22}
A.~Ghosh, S.~A. Pawar, and R.~I. Sujith.
\newblock Anticipating synchrony in dynamical systems using information theory.
\newblock {\em Chaos}, 32:031103, 2022.

\bibitem{romano04}
M.~C. Romano, M.~Thiel, and J.~Kurths.
\newblock Generalized synchronization indices based on recurrence in phase
  space.
\newblock {\em AIP Conf. Proc.}, 742:330, 2004.

\bibitem{romano05}
M.~C. Romano, M.~Thiel, J.~Kurths, I.~Z. Kiss, and J.~L. Hudson.
\newblock Detection of synchronization for non-phase-coherent and
  non-stationary data.
\newblock {\em Europhys. Lett.}, 71:466, 2005.

\bibitem{ghosh20_2}
A.~Ghosh and R.~I. Sujith.
\newblock Emergence of order from chaos: {A} phenomenological model of coupled
  oscillators.
\newblock {\em Chaos Solitons Fractals}, 141:110334, 2020.

\bibitem{hampton99}
A.~Hampton and D.~H. Zanette.
\newblock Measure synchronization in coupled {H}amiltonian systems.
\newblock {\em Phys. Rev. Lett.}, 83:2179, 1999.

\bibitem{wang03}
X.~Wang, M.~Zhan, C.-H. Lai, and H.~Gang.
\newblock Measure synchronization in coupled ${\ensuremath{\varphi}}^{4}$
  {H}amiltonian systems.
\newblock {\em Phys. Rev. E}, 67:066215, 2003.

\bibitem{gupta19}
D.~S. Gupta and A.~Bahmer.
\newblock Increase in mutual information during interaction with the
  environment contributes to perception.
\newblock {\em Entropy}, 21:365, 2019.

\bibitem{ameri15}
V.~Ameri, M.~Eghbali-Arani, A.~Mari, A.~Farace, F.~Kheirandish, V.~Giovannetti,
  and R.~Fazio.
\newblock Mutual information as an order parameter for quantum synchronization.
\newblock {\em Phys. Rev. A}, 91:012301, 2015.

\bibitem{wilmer12}
A.~Wilmer, M.~de~Lussanet, and M.~Lappe.
\newblock Time-delayed mutual information of the phase as a measure of
  functional connectivity.
\newblock {\em PLOS ONE}, 7:e44633, 2012.

\bibitem{fan21}
H.~Fan, L.~Kong, Y.~Lai, and X.~Wang.
\newblock Anticipating synchronization with machine learning.
\newblock {\em Phys. Rev. Res.}, 3:023237, 2021.

\bibitem{mondal17_chaos}
S.~Mondal, S.~A. Pawar, and R.~I. Sujith.
\newblock Synchronous behaviour of two interacting oscillatory systems
  undergoing quasiperiodic route to chaos.
\newblock {\em Chaos}, 27:103119, 2017.

\bibitem{raaj19}
A.~Raaj, J.~Venkatramani, and S.~Mondal.
\newblock Synchronization of pitch and plunge motions during intermittency
  route to aeroelastic flutter.
\newblock {\em Chaos}, 29:043129, 2019.

\bibitem{yu01}
P.~Yu and A.B. Gumel.
\newblock Bifurcation and stability analyses for a coupled {B}russelator model.
\newblock {\em J. Sound Vib.}, 244:795, 2001.

\bibitem{nurujjaman06}
M.~D. Nurujjaman and A.~N. Sekar~Iyengar.
\newblock Chaotic-to-ordered state transition of cathode-sheath instabilities
  in {DC} glow discharge plasmas.
\newblock {\em Pramana-J Phys.}, 67:299, 2006.

\bibitem{nurujjaman07}
M.~Nurujjaman, R.~Narayanan, and A.~N. Sekar~Iyengar.
\newblock Parametric investigation of nonlinear fluctuations in a dc glow
  discharge plasma.
\newblock {\em Chaos}, 17:043121, 2007.

\bibitem{nguyen13}
L.~H. Nguyen and K.~Hong.
\newblock Adaptive synchronization of two coupled chaotic {H}indmarsh-{R}ose
  neurons by controlling the membrane potential of a slave neuron.
\newblock {\em Appl. Math. Model.}, 37:2460, 2013.

\bibitem{seshadri16}
A.~Seshadri and R.~I. Sujith.
\newblock A bifurcation giving birth to order in an impulsively driven complex
  system.
\newblock {\em Chaos}, 26:083103, 2016.

\bibitem{lorenz63}
E.~N. Lorenz.
\newblock Deterministic nonperiodic flow.
\newblock {\em J. Atmos. Sci.}, 20:130, 1963.

\bibitem{roessler76}
O.~E. R\"ossler.
\newblock An equation for continuous chaos.
\newblock {\em Phys. Lett. A}, 57:397, 1976.

\bibitem{chen99}
G.~Chen and T.~Ueta.
\newblock Yet another chaotic attractor.
\newblock {\em Int. J. Bifurc. Chaos}, 09:1465, 1999.

\bibitem{mainieri99}
R.~Mainieri and J.~Rehacek.
\newblock Projective synchronization in three-dimensional chaotic systems.
\newblock {\em Phys. Rev. Lett.}, 82:3042, 1999.

\bibitem{cover06}
T.~M. Cover and J.~A. Thomas.
\newblock {\em Elements of information theory}.
\newblock Wiley-Interscience Press, USA, second edition, 2006.

\bibitem{sprott03}
J.~C. Sprott.
\newblock {\em Chaos and Time-Series Analysis}.
\newblock Oxford University Press, New York, first edition, 2003.

\bibitem{abarbanel96}
H.~D.~I. Abarbanel, N.~F. Rulkov, and M.~M. Sushchik.
\newblock Generalized synchronization of chaos: The auxiliary system approach.
\newblock {\em Phys. Rev. E}, 53:4528, 1996.

\end{thebibliography}
\end{document}